 \documentclass[final,5p,times,twocolumn]{elsarticle}

\usepackage{subfigure}
\usepackage{amssymb}
\usepackage{lipsum}
\usepackage{amsthm}
\usepackage{graphicx}
\usepackage{amsmath}
\usepackage[colorlinks=true, linkcolor=blue, citecolor=blue, urlcolor=blue]{hyperref}
\usepackage{subcaption}

\journal{Physics of the Dark Universe}

\begin{document}

\begin{frontmatter}



    \title{Chaotic Imprints in Gravitational Waves from Conformal-Anomaly-Corrected Extreme-Mass-Ratio Inspirals }


\author[first]{Wei-Hao Zhang}
\ead{zhangweihao0909@163.com}
\author[first,second]{Yu-Sen An\fnref{label1}}
\fntext[label1]{Corresponding author}
\ead{anyusen@nuaa.edu.cn}
\affiliation[first]{organization={Center for the Cross-disciplinary Research of Space Science and Quantum-technologies (CROSS-Q), College of Physics, Nanjing University of Aeronautics and Astronautics, Nanjing, 210016, China}}
\affiliation[second]{organization={MIIT Key Laboratory of Aerospace Information Materials and Physics,  Nanjing University of Aeronautics and Astronautics, Nanjing, 210016, China}}

\begin{abstract}
In this work, we investigate the effect of chaotic orbits on the extreme mass ratio inspiral (EMRI) gravitational wave signals where the central black hole is corrected by quantum conformal anomaly. 
We utilize the numerical kludge method to compute gravitational waveforms produced by the compact object along different orbital trajectories, and also derive the corresponding frequency distribution and energy spectra of gravitational waves. Our calculations reveal that variations in orbital energy or anomaly coefficient drive the orbital evolution from regular integrable motion to chaotic motion, and such dynamical transition leaves clear imprints on gravitational-wave signal. Specifically, gravitational waves originating from chaotic orbits feature pronounced irregular and time-varying amplitude fluctuations, accompanied by abundant fine spectral spikes and extended continuous spectral distributions in both frequency and energy domains, which differ drastically from the gravitational radiation generated by the regular non-chaotic orbits. Moreover, we evaluate the detectability by comparing the calculated characteristic strain of gravitational waves emitted by the compact object on different orbits with the sensitivity curves of future space-based GW detectors, including LISA, Taiji and TianQin. The results demonstrate that these detectors are capable of capturing gravitational-wave signals from chaotic systems modified by conformal anomalies, which provide a potential pathway for detecting conformal anomaly correction in astronomical observation.
\end{abstract}



\begin{keyword}
black hole \sep chaos \sep geodesic \sep conformal anomaly \sep gravitational waves



\end{keyword}

\end{frontmatter}




\section{Introduction:}
General Relativity (GR) is currently a fundamental and highly effective theoretical framework for describing gravitational phenomena. The validity of this fundamental theory has been validated by abundant observations covering multiple distinct scales ranging from weak-field regime such as Solar System to strong-field regimes such as black holes \cite{Peters:1964zz,Einstein:1937qu}. 
These observations include the detection of gravitational waves (GW) accomplished by the LIGO and Virgo scientific collaborations \cite{LIGOScientific:2016aoc,LIGOScientific:2019zcs,LIGOScientific:2016emj}, along with the imaging observations of supermassive black holes carried out by the Event Horizon Telescope (EHT) \cite{EventHorizonTelescope:2019uob,EventHorizonTelescope:2022wkp}. Such observational results further verify that black holes predicted by general relativity are physically real astrophysical objects instead of pure theoretical solutions. Particularly remarkable is the landmark detection of gravitational waves from the LIGO and Virgo teams. This breakthrough not only confirms the physical existence of black holes foretold by general relativity, but also signals the birth of gravitational-wave astronomy, offering a brand-new research avenue for our future astronomical investigations. 

It is now widely accepted that the spacetime in the near-horizon region of black holes undergoes extreme curvature, which endows particles surrounding black holes with distinctive dynamical behaviours in such environments. For instance, massless particles such as photons will be captured by black holes when they approach the region of the black hole event horizon. This phenomenon gives rise to the formation of black hole shadows,an effect that has received direct observational confirmation \cite{EventHorizonTelescope:2019uob,EventHorizonTelescope:2022wkp,EventHorizonTelescope:2022apq}.

Furthermore, massive charged particles are capable of undergoing oscillatory motions along stable circular orbits, generating high-frequency quasi-periodic oscillation signals accessible to observation \cite{Ozel:2010bz}. These observational findings concerning particle motions adjacent to black holes provide an extremely vital research pathway for investigating and interpreting the geometric structure of black holes as well as the underlying gravitational theories. 

Among various phenomena of particle motion around black holes, the chaotic motion of particles in the near-horizon region of black holes has attracted widespread attention from the academic community, and numerous studies have investigated this phenomenon from diverse perspectives. It is pointed out in Ref.\cite{Hashimoto:2016dfz,Bera:2021lgw,Dalui:2018qqv} that the event horizon of a black hole can trigger chaotic particle motion. If chaotic phenomena are investigated within the framework of classical gravity, two classic chaos characterization indicators, namely the Poincaré section and Lyapunov exponent can be used for description. Existing studies \cite{Hashimoto:2016dfz} have found that the Lyapunov exponent of chaotic particle motion near the Schwarzschild black hole saturates the universal upper bound found in \cite{Maldacena:2015waa} where $\lambda_L\le\kappa$. Ref.\cite{Yu:2023spr,Dutta:2024rta,Gao:2022ybw,Gwak:2022xje,Lei:2020clg} also investigate this chaotic bound and potential bound-violation phenomena. Ref.\cite{Xie:2025auj,Zhang:2025cdx,Shukla:2024tkw,Guo:2022kio} investigate the correlation between chaos and black hole phase transitions.While Ref.\cite{Bera:2021lgw,Das:2024iuf} focused on particle chaotic motion in modified gravity theories with quantum corrections. It should be noted that, apart from the research paradigm of chaos via classical methods, the chaotic properties of black holes can also be studied from a holographic perspective, that is, characterizing chaotic phenomena through the out-of-time-order correlator (OTOC) of boundary systems \cite{Roberts:2014isa,Shenker:2013pqa}. 

As is well known, GR falls into the category of classical gravitational theories, in which no quantum correction effects are taken into account. Accordingly, all dynamical and thermodynamic features predicted by this theory purely possess classical properties. To examine the impacts on the theory after incorporating quantum corrections, one of the existing theoretical frameworks directly introduces quantum field theory in the curved spacetime background. Under such circumstances, the classical stress-energy tensor ought to be replaced by the expectation value of the stress-energy operator, and thus the resulting equation should be 
\begin{equation}\label{Einstein}
    R_{\mu\nu}-\frac{1}{2}g_{\mu\nu}R=8\pi G \langle T_{\mu\nu}\rangle
\end{equation}
For the vacuum configuration in flat spacetime, expectation value $\langle T_{\mu\nu}\rangle$ should be zero.\footnote{Note that in curved spacetime, there exists various kinds of choices of vacuum state, here we choose Hartle Hawking vaccum since we want to keep the horizon smooth.} However, for curved spacetime, there is conformal anomaly which leads to non-vanishing trace of stress energy tensor \cite{Deser:1993yx}. Ref.\cite{Cai:2009ua} investigate the back-reaction of this conformal anomaly on the black hole geometry where the authors derived an analytical black hole solution. This spacetime is a good platform to investigate the quantum corrections on black hole physics. While Ref.\cite{Cai:2009ua} only considers the static and spherical symmetric black hole in asymptotically flat spacetime, there are further generalizations to asymptotic AdS case \cite{Cai:2014jea}, rotating case \cite{Fernandes:2023vux} and Vaidya case \cite{Gurses:2023ahu}. 

Existing research has demonstrated that the thermodynamic properties of black hole spacetimes can be significantly modified once quantum anomaly corrections are incorporated. Such modifications show most prominently in black hole entropy, which no longer obeys the Bekenstein area law but acquires a logarithmic correction term $S=\frac{A}{4G}+\alpha ln\frac{A}{A_0}$. Furthermore, this logarithmic correction shows universal behavior across current quantum gravity frameworks \cite{Solodukhin:1997yy,Kaul:2000kf,Das:2001ic}. Apart from this, AdS black holes supplemented with conformal anomaly corrections experience dramatic shifts in their phase transition behaviors. One noteworthy feature arising in this context deserves detailed discussion: the emergence of an "isolated critical point" within the black hole phase diagram \cite{Dolan:2014vba}. The critical exponents associated with this second-order transition lie outside the predictive scope of mean-field theory and violate the Rushbrooke scaling law \cite{Hu:2024ldp}. 

It can be observed that conformal anomalies exert numerous pronounced influences on black hole thermodynamics, yet their underlying mechanism governing particle dynamics remains unclear, and systematic in-depth investigations within this research domain are still lacking. Ref.\cite{Zhang:2023bzv} explored the shadow images generated by black holes with conformal anomaly corrections, but this work solely considered regular particle motion around black holes without addressing the chaotic particle motion discussed above—a typical motion that is highly likely to show distinctive observational signatures. 

To probe this possibility, we preliminarily investigated the chaotic dynamical properties of particles surrounding asymptotically flat black holes subject to quantum corrections in our earlier work \cite{An:2025xmb}. Two kinds of particle motions were constructed in that study. The first model describes massive neutral particles confined to the  $r-\theta$ plane, an external background harmonic potential is introduced to stabilize particle orbits, enabling an analysis of how the presence of the black hole event horizon modifies the dynamical behaviors of integrable particles, with Poincaré sections adopted as the diagnostic tool for identifying chaos. The second model concerns charged particles undergoing radial motion. In this setup, we focus on massive charged particles moving purely along the radial coordinate with an electrostatic potential included, to examine the dynamical evolution of particles under the given field configuration, and characterize chaotic behaviors via the calculation of Lyapunov exponents. Our core research conclusion is as follows: quantum conformal anomaly corrections significantly amplify the chaotic motion of particles near black hole event horizons and reshape the conventional patterns of dynamical evolution. Such amplification effect becomes increasingly prominent as the anomaly coefficient rises, and this result receives consistent validation across both particle models and both chaos characterization approaches. 

The present work continues to center its research on the chaotic motion of particles residing in the spacetime of black holes modified by quantum conformal anomalies. Distinct from our previous studies that qualitatively investigated the theoretical impacts of conformal anomaly corrections on particle dynamical behaviors, this paper carries out quantitative research on chaos by using GW signals to characterize the chaotic motion of particles. An EMRI system generally arises from an inspiral system composed of stellar-mass compact objects (with masses ranging from $10^5-10^{10}$ solar masses) revolving around a supermassive black hole (SMBH, whose mass spans $10^0-10^1$ solar masses), and such celestial objects are commonly distributed in the nuclear region of the Milky Way Galaxy. As the mass ratio is huge, the stellar-mass compact object can be approximated as a particle so the chaotic particle motion will directly affect the gravitational wave signals for such EMRI systems. Recently Ref.\cite{Das:2025eiv} considered the chaotic dynamical phenomena within an Extreme Mass Ratio Inspiral (EMRI) configuration whose configuration corresponds to a stellar-mass compact object orbiting close to the event horizon of a supermassive Schwarzschild black hole embedded within a Dehnen-(1,4,5/2) dark matter halo.  

As mentioned earlier, the historic detection of gravitational waves by the LIGO and Virgo collaborations has ushered in a new era for humanity’s future astronomical observations. A variety of systems, including binary white dwarf systems     \cite{Huang:2020rjf,Korol:2017qcx}, massive binary black hole systems \cite{Klein:2015hvg,Wang:2019ryf}, stellar-mass binary black hole systems \cite{Sesana:2016ljz,Liu:2020eko} and extreme mass ratio inspiral systems \cite{Babak:2017tow,Fan:2020zhy}, are predicted to emit gravitational wave signals predominantly within the millihertz (mHz) frequency band. Space-based gravitational wave detectors are specially engineered to capture gravitational wave signals emitted by these sources at millihertz frequencies \cite{TianQin:2020hid,LISA:2017pwj,Hu:2017mde}. Among all candidate target sources, EMRI systems stand out as particularly vital, since their signals can persist for thousands up to millions of orbital cycles, which grants a unique opportunity to conduct precision tests of GR under strong-field regimes \cite{Yunes:2011aa,Canizares:2012is,Moore:2017lxy}. It is well established that chaotic orbits can alter the gravitational wave signals generated by EMRI systems \cite{Das:2025eiv}.Based on such intrinsic correlation,it is reasonable to conjecture that the emitted GW signals contain hidden information pertaining to chaotic orbits modulated by conformal anomalies.Since these signals can be directly measured by future space-based GW detectors such as LISA,Taiji and TianQin,this opens up a direct avenue for observational searches of conformal anomaly effects in astrophysical sources.These existing research advances and the broad outlook of this research field lay solid foundations and provide guidance for the present investigation. 

Based on our previous work \cite{An:2025xmb},the present study adopts the numerical kludge method to generate gravitational wave waveforms radiated by EMRI configurations embedded in the spacetime of spherically symmetric black holes corrected by quantum conformal anomalies,with a central focus on examining how chaos amplified by quantum conformal anomalies alters gravitational wave signals. We systematically compare the gravitational wave waveforms, frequency spectra and energy spectra produced by non-chaotic orbits,marginally chaotic orbits at the onset of chaos,and fully chaotic orbits modified by different values of the conformal anomaly coefficient. Meanwhile,we analyze the observational signatures of conformal-anomaly-induced chaos encoded within gravitational wave signals. One key result of this study reveals that chaotic phenomena amplified by conformal anomalies can be detected 
by future space-based gravitational wave detectors including LISA, Taiji and TianQin. Our results lay a theoretical foundation for the practical observational search of conformal anomaly effects in future astronomical investigations. 

The structure of this paper is as follows.In Sec.\ref{sec2}, we briefly review the dynamical properties of EMRI systems residing in the analytic black hole solutions with conformal anomaly corrections. Afterwards, Sec.\ref{sec3} introduces the numerical kludge method to calculate gravitational wave waveforms, followed by an in-depth analysis of their characteristic modes, average amplitudes, radiative power, frequency spectra and energy spectra.
Finally,we adopt a simple and efficient method to assess the detection capability of future space-based GW detectors (including LISA, TianQin and Taiji) for chaotic gravitational-wave signals. The findings obtained in this work are summarized in Sec.\ref{sec4}.
Throughout this paper, the natural unit system is used where $G=1$,$c=1$ and $\hbar=1$. 
\section{Black hole with conformal anomaly correction:}\label{sec2}
In this section, we first review the analytical solution of the black hole with quantum conformal anomaly corrections, which was used in our previous work. We consider the following simplification: a static, spherically symmetric spacetime  \cite{Cai:2009ua}
\begin{equation}\label{metric}
    ds^{2}=-f(r)dt^{2}+\frac{1}{f(r)}dr^{2}+r^{2}(d\theta^{2}+\sin^{2}\theta d\phi^{2})
\end{equation}
where
\begin{equation}\label{blacken}
    f(r)=1-\frac{r^{2}}{4\alpha}(1-\sqrt{1-8\alpha(\frac{2M}{r^{3}}-\frac{Q^{2}}{r^{4}})})
\end{equation}
where $\alpha=8\pi\tilde{\alpha}$ denotes the renormalized central charge(conformal anomaly coefficient), which takes a positive real value. 
Here,$M$ stands for the ADM mass of the black hole, and $Q$ represents the $U(1)$ charge associated with the underlying conformal field theory, this charge cannot be simply set to zero as discussed in Ref.\cite{Cai:2009ua,Cai:2014jea}. The event horizon radius of the black hole corresponds to the largest root of the equation $f(r_h)=0$, which reads 
\begin{equation}
    r_{h}=M+\sqrt{M^{2}-Q^{2}+2\alpha}
\end{equation}
From this, the mass and charge should satisfy the following relation in order to avoid the naked singularity
\begin{equation}
    M^{2}\geqslant Q^{2}-2\alpha. 
\end{equation}
Now we  review the dynamical behaviors of massive particles moving in the spacetime of black holes modified by quantum conformal anomalies. For the convenience of subsequent analysis, we adopt Painlevé coordinates. This coordinate system is chosen because it remains well-defined at the event horizon and enables more robust numerical calculations of particle trajectories in the near-horizon region \cite{Martel:2000rn,Baumgarte:2010ndz}. After coordinate transformation, Eq.(\ref{metric})takes the following form:\begin{equation}
    ds^2=-f(r)dt^2+2\sqrt{1-f(r)}dtdr+dr^2+r^2(d\theta^2+sin^2\theta d\phi^2)
\end{equation}
The four momentum of the massive particles satisfy the following relation 
\begin{equation}    g^{\mu\nu}p_{\mu}p_{\nu}=-p_{t}^{2}+2\sqrt{1-f(r)} p_{r}p_{t}+(f(r)p_{r}^{2}+\frac{p_{\phi}^{2}}{r^{2}})=-m^{2}
\end{equation}
In this work, we adopt a simplified model with equatorial plane symmetry (i.e.$\theta=\frac{\pi}{2}$), which yields $p_\theta=0$.\footnote{In the previous work \cite{An:2025xmb}, we investigate the $r-\theta$ plane dynamics, however the equations of motion are the same as $r-\phi$ plane when $\theta=\frac{\pi}{2}$, so the resulting Poincare sections are the same as in Ref.\cite{An:2025xmb}. } Consequently, the conserved energy of outward-emitted particles satisfies the following relation: \begin{equation}
    E=-\sqrt{1-f(r)}p_{r}+\sqrt{p_{r}^{2}+\frac{p_{\phi}^{2}}{r^{2}}+m^{2}}
\end{equation}
Likewise, to observe the transition of particle motion from integrable dynamics to chaos near the black hole over an extended evolution time, we introduce a simple external harmonic oscillator potential to confine the particles. Under this setup, the total energy of the particle reads\begin{align}\label{energy}
    E=&-\sqrt{1-f(r)}p_{r}+\sqrt{p_{r}^{2}+\frac{p_{\phi}^{2}}{r^{2}}+m^{2}}+\frac{1}{2}K_{r}(r-r_{c})^{2}\notag \\
   & +\frac{1}{2}K_{\phi} r_{h}^{2}(\phi-\phi_{c})^{2}
\end{align}
where $r_c$,$\phi_c$is the balance point. Accordingly, one can derive the following set of dynamical equations of motion:\begin{equation}
    \dot{r}=-\sqrt{1-f(r)}+\frac{p_{r}}{\sqrt{p_{r}^{2}+\frac{p_{\phi}^{2}}{r^{2}}+m^{2}}}
\end{equation}
\begin{equation}
    \dot{p}_{r}=-\frac{f'(r)}{2\sqrt{1-f(r)}} p_{r}+\frac{p_{\phi}^{2}/r^{3}}{\sqrt{p_{r}^{2}+p_{\phi}^{2}/r^{2}+m^{2}}}-K_{r}(r-r_{c})
\end{equation}
\begin{equation}
    \dot{\phi}=\frac{p_{\phi}/r^{2}}{\sqrt{p_{r}^{2}+p_{\phi}^{2}/r^{2}+m^{2}}}
\end{equation}
\begin{equation}
\dot{p}_{\phi}=-K_{\phi} r_{h}^{2}(\phi-\phi_{c})
\end{equation}

\begin{figure*}[h]
    \centering
    \subfigure[$\alpha=0.01$]{\includegraphics[width=0.33\textwidth]{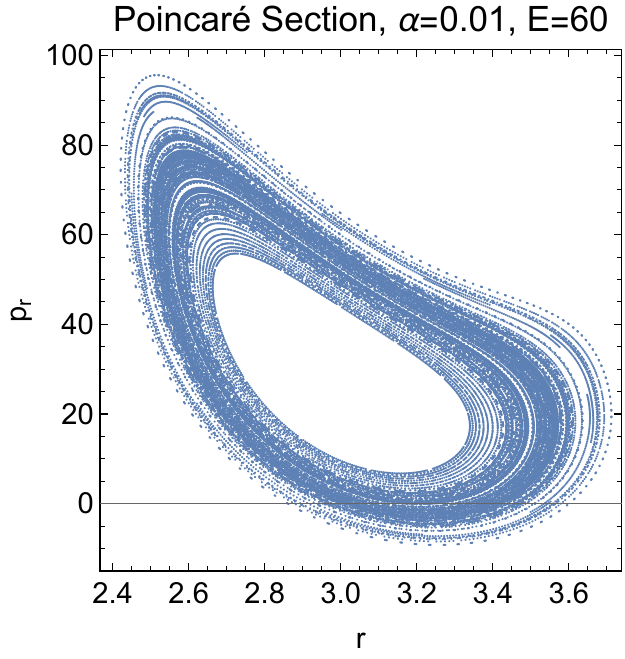}}\hfill
    \subfigure[$\alpha=0.1$]{\includegraphics[width=0.33\textwidth]{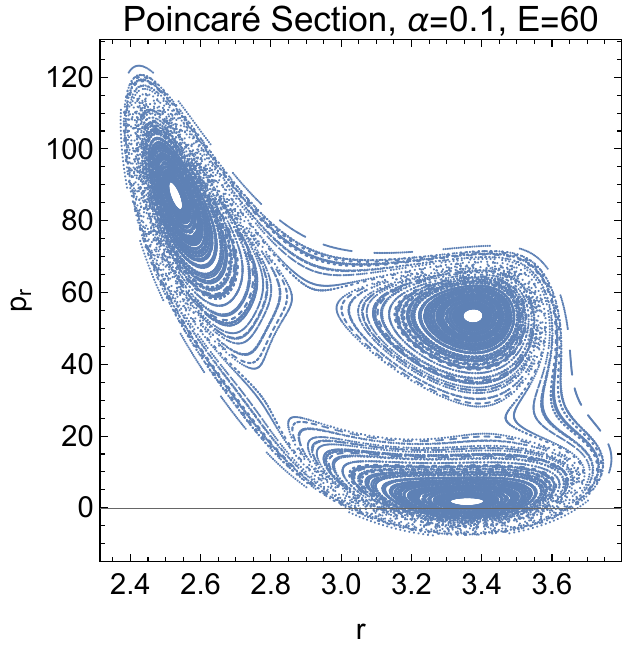}}\hfill
    \subfigure[$\alpha=0.25$]{\includegraphics[width=0.33\textwidth]{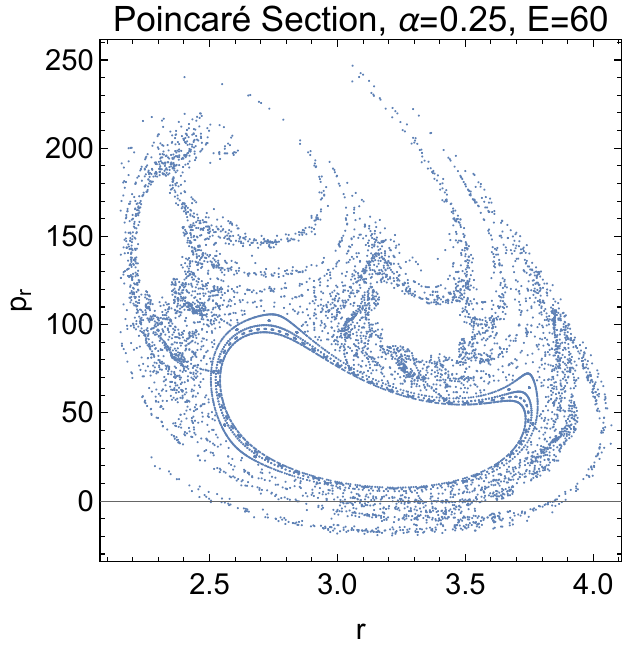}}\hfill
    \caption{The change of Poincare section of massive particles for $E = 60$ and different $\alpha$, as anomaly $\alpha$ increase, the Poincare section becomes distorted and irregular which means that the particle motion becomes more chaotic.This behavior is previously found in our work \cite{An:2025xmb}.}
    \label{psm1}
\end{figure*}

In our prior study \cite{An:2025xmb}, after deriving the above equations of particle motion, we assign simple parameter values for subsequent numerical calculations: the $U(1)$ charge is set to be $Q=\frac{1}{\sqrt{2}}$, the harmonic oscillator potential parameters $K_r=80$, $K_\phi=20$, $r_c=3.2$ and $\phi_c=0$. 
then we fix the total energy of the neutral massive particle at $E=60$ and vary the anomaly coefficient as $\alpha=0.01, 0.1, 0.25$. 
We then plot Poincaré sections for all parameter combinations to observe the dynamical transition of particle motion from integrable to chaotic behaviors, as illustrated in Fig.\ref{psm1}. We find that as the anomaly coefficient increases, the Poincaré sections evolve from regular smooth tori to distorted surfaces with minor fractures, and eventually become fragmented. This evolution corresponds to a shift of particle orbits from non-chaotic orbits to onset-of-chaos orbits, and finally to the largely chaotic orbits. These qualitative results confirm that conformal anomaly effects do modify the chaotic motion of test particles. 

Based on these findings, the remainder of this paper focuses on gravitational-wave signatures radiated by chaotic orbits within EMRI systems embedded in this black hole spacetime. We note that the main body of this work centers on how different quantum conformal anomaly coefficients affect gravitational radiation, while the influence of varying particle energies \footnote{Increasing particle's energy with fixed anomaly parameter can also induce chaotic particle motions.} on system dynamics is discussed separately in the appendix. 

\section{Numerical Kludge gravitational waveforms from EMRIs:} \label{sec3}
In this section, we describe the procedures for calculating gravitational waveforms generated by a steller-mass celestial objects orbiting black holes with quantum conformal anomaly corrections. The astrophysical system under consideration here is the EMRI system with a primary central object and a secondary orbiting object. The primary component is typically a supermassive black hole, while compact stars with much lower masses serve as secondary objects and we choose the mass ratio to be $m/M=10^{-5}$ without loss of generality.  

Strictly speaking, these gravitational waves radiate the energy and angular momentum of the secondary object outwards, driving the secondary body to follow an inward inspiral trajectory. Nevertheless, over short evolution timescales, the energy and angular momentum losed by the secondary object are negligible compared with the total energy of the entire system, which supports the adiabatic approximation herein \cite{Grossman:2011im,Zi:2023qfk,Sundararajan:2007jg}. This approximation states that if the inspiral timescale of the small secondary body is far longer than its orbital period, the loss of energy and angular momentum induced by outward gravitational wave radiation proceeds at an extremely slow rate and can be neglected over dozens of orbital cycles. Moreover, any backreaction effects such as the influence of gravitational radiation on the motion of the secondary object can also be disregarded. Under this approximation, the motion of the low-mass secondary object can be characterized as geodesic trajectories within the spacetime of the supermassive black hole. The resulting gravitational wave signals therefore encode information about both chaotic and non-chaotic orbits, and naturally contain imprints of how conformal anomaly corrections modify chaotic dynamics, a conclusion drawn from our prvious work. 

The numerical kludge method is adopted herein to compute the GW waveforms produced by the orbital motion of the secondary compact object around the central supermassive black hole \cite{Babak:2006uv}. To calculate the waveforms via this method, the orbital trajectory of the secondary object must first be acquired. Specifically, we numerically integrate the dynamical equations of the secondary particle derived under Painlevé coordinates in Sec.\ref{sec2} of this paper along the orbital path, which yields the coordinates of the secondary object as functions of time. Afterwards, the obtained time-dependent orbital trajectory of the secondary body is projected onto a spherical polar coordinate grid of flat spacetime to construct an equivalent trajectory in Minkowski spacetime, where the corresponding coordinate system is the Cartesian coordinate system on a Euclidean plane, expressed as follows: \begin{align}
    x(t)&=r(t)\sin\theta(t)\cos\phi(t),\notag\\
    y(t)&=r(t)\sin\theta(t)\sin\phi(t),\notag\\
    z(t)&=r(t)\cos\theta(t),
\end{align}
As stated in Ref.\cite{Babak:2006uv}, the projected trajectory of the secondary object in flat spacetime merely serves as the input for phenomenological waveform generation modeling. Despite the 
gravitational wave amplitudes derived from numerical kludge method, the extracted gravitational wave frequency distribution which remains accurate even under approximation is also a very important quantity to encode the physical information. Accordingly, after establishing the orbital trajectory of the secondary object, these reconstructed waveforms and frequency distribution can be utilized for data analysis. 

With the orbital trajectory of the secondary object fully determined, we then substitute this trajectory into the gravitational quadrupole radiation formula to generate the corresponding gravitational waveforms. In the weak-field regime, the spacetime metric takes the form $g_{\mu\nu}=\eta_{\mu\nu}+h_{\mu\nu}$, where $\eta_{\mu\nu}$ denotes the Minkowski metric and $h_{\mu\nu}$ stands for the tiny metric perturbation satisfying $h_{\mu\nu}\ll1$. The trace-reversed metric perturbation is defined via the relation $\bar{h}^{\mu\nu}=h^{\mu\nu}-\frac{1}{2}\eta^{\mu\nu}h$, in which $h=\eta^{\mu\nu}h_{\mu\nu}$ corresponds to the trace of the perturbation. Imposing the Lorenz gauge condition $\partial_\nu h^{\mu\nu}=0$, one can derive the linearized Einstein field equations: \begin{equation}\label{wave equation}
    \Box \bar{h}^{\mu\nu}=-16\pi\mathcal{T} ^{\mu\nu}
\end{equation}
where $ \Box$ denotes the standard d'Alembertian operator in flat spacetime, and $\mathcal{T} ^{\mu\nu}$ refers to the effective energy-momentum tensor that obeys the conservation relation $\partial_\nu \mathcal{T} ^{\mu\nu}=0$. In the coordinate system centered on the black hole, a particular solution to the above wave equation Eq.(\ref{wave equation})  can be obtained, which takes the familiar form of retarded potentials: \begin{equation}\label{retarded potential}
    h_{\mu\nu}(\textbf{x},t)=4G\int d^3\textbf{x}'\frac{\mathcal{T} ^{\mu\nu}(\textbf{x}',t-|\textbf{x}-\textbf{x}'|)}{\textbf{x}-\textbf{x}'}
\end{equation}
where,the observer coordinate and source coordinate are denoted by $\textbf{x}'$ and $\textbf{x}$ respectively, and this expression describes the gravitational radiation generated by the source term $\mathcal{T} ^{\mu\nu} $. The quantity $t'=t-|\textbf{x}-\textbf{x}'|$ is defined as retarded time, which means the gravitational field at position $\textbf{x}$ and time $t$ is determined by the source at position $\textbf{x}$' evaluated at the earlier retarded time $t'$. When the motion of the source is only weakly modified by gravitational effects, $\mathcal{T} ^{\mu\nu}$ in Eq.(\ref{retarded potential}) can be approximated by the matter energy-momentum tensor ${T} ^{\mu\nu}$ \cite{Babak:2006uv}. In the slow-motion limit, Eq.(\ref{retarded potential})  reduces to the standard quadrupole formula \cite{Babak:2006uv} 
\begin{equation}\label{quadrupole formula}
    \bar{h}^{ij}=\frac{2G}{r}\Big{[} \ddot{I}^{ij}(t')\Big{]}_{t'=t-|\textbf{x}-\textbf{x}'|}
\end{equation}
where\begin{equation}\label{quadrupole moment}
   {I}^{ij}=\int x^i x^j {T} ^{00}(t,x^i)d^3x
\end{equation}
stands for the symmetric and traceless mass quadrupole moment of the source.For a small-mass secondary compact object with mass $m$ whose orbital trajectory is described by $Z^i(t)$, the $tt$-component of the energy-momentum tensor is given by $T^{00}$, which is defined as \cite{Thorne:1980ru}
\begin{equation}
    T^{tt} (t,x^i)= m\delta^3({x^{i}} - {Z^i}(t))
\end{equation}

Upon inserting Eq.(\ref{quadrupole moment}) into Eq.(\ref{quadrupole formula}),we arrive at the gravitational-wave quadrupole expression in the slow-rotation approximation \cite{Press:1977ps,Babak:2006uv,Poisson_Will_2014}:\begin{equation}\label{simplify quadrupole formula}
    h_{ij}=\frac{2G}{D_L}\frac{d^2I_{ij}}{dt^2}=\frac{2Gm}{D_L}(a_ix_j+a_jx_i+2v_iv_j)
\end{equation}
In this context, ${D}_L$ is the luminosity distance to the detector, $ v_i$ and $ a_i$ are spatial velocity and acceleration of the secondary stellar-massive compact object, respectively.

To investigate gravitational wave signals received by detectors, we construct a detector-adapted coordinate system ($X$,$Y$,$Z$) \cite{Poisson_Will_2014,Yang:2024lmj}. This coordinate system shares the same origin as the original ($x$,$y$,$z$) coordinate system, with the supermassive black hole placed at the origin for both frames. We define the basis vectors $(\hat{e}_X,\hat{e}_Y, \hat{e}_Z)$ of the detector-adapted coordinate system. The coordinate transformation between this new frame and the original frame is governed by two parameters: the inclination angle $\iota$ between the orbital plane and the $X-Y$ plane, and the longitude of periapsis $\zeta$ within the orbital plane, as expressed below \cite{Poisson_Will_2014} : \begin{align}
\hat{e}_{X}&=(\cos\,\zeta,-\sin\,\zeta,0),\\
\hat{e}_{Y}&=(\sin\,\iota \sin\,\zeta,\cos\,\iota \cos\,\zeta,-\sin\,\iota),\\
\hat{e}_{Z}&=(\sin\,\iota \sin\,\zeta,-\sin\,\iota \cos\,\zeta,\cos\,\iota),
\end{align}
The GW polarizations can then be derived by projecting the metric perturbations Eq.(\ref{simplify quadrupole formula}) onto the detector frame,yielding:\begin{align}
    h_{+}&=\frac{1}{2}(\hat{e}^i_{X} \hat{e}^j_{X}-\hat{e}^i_{Y} \hat{e}^j_{Y})h_{ij},\\
    h_{\times}&=\frac{1}{2}(\hat{e}^i_{X} \hat{e}^j_{Y}-\hat{e}^i_{Y} \hat{e}^j_{X})h_{ij},
\end{align}
The above polarization components can be expressed as specific linear combinations of the components $h_{\zeta\zeta}$, $h_{\iota\iota}$,$h_{\iota\zeta}$ ,as detailed below: \begin{align}
    &h_{+}=\frac{1}{2}(h_{\zeta\zeta}-h_{\iota\iota}),\\
    &h_{\times}=h_{\iota\zeta},
\end{align}
in which the corresponding components are given by \cite{Babak:2006uv}
\begin{align}
    h_{\zeta\zeta}=&\; h_{xx}\cos^2\,\zeta-h_{xy}\sin2\,\zeta+h_{yy}\sin^2\,\zeta,\\
    h_{\iota\iota}=& \; \cos^2\,\iota[h_{xx}\sin^2\,\zeta+h_{xy}\sin2\,\zeta+h_{yy}\cos^2\,\zeta]\notag \\
    &+h_{zz}\sin^2\,\iota-\sin2\,\iota[h_{xz}\sin\,\zeta+h_{yz}\cos\,\zeta],\\
h_{\iota\zeta}=&\; \frac{1}{2}\cos\,\iota[h_{xx}\sin2\,\zeta+2h_{xy}\cos2\,\zeta-h_{yy}\sin2\,\zeta]\notag \\
&+\sin\,\iota[h_{yz}\sin\,\zeta-h_{xz}\cos\,\zeta],
\end{align}
Subsequently,we will adopt the methods elaborated in this section to conduct our gravitational wave analysis. 
\begin{figure*}[ht]
    \centering
    \subfigure[]{\includegraphics[width=0.45\textwidth]{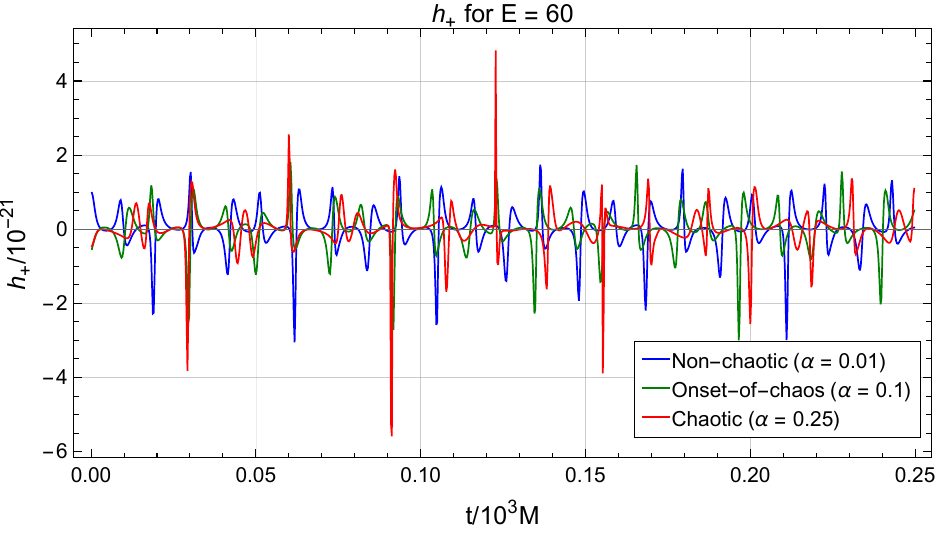}\label{Fig.2(a)}}\hfill
    \subfigure[]{\includegraphics[width=0.45\textwidth]{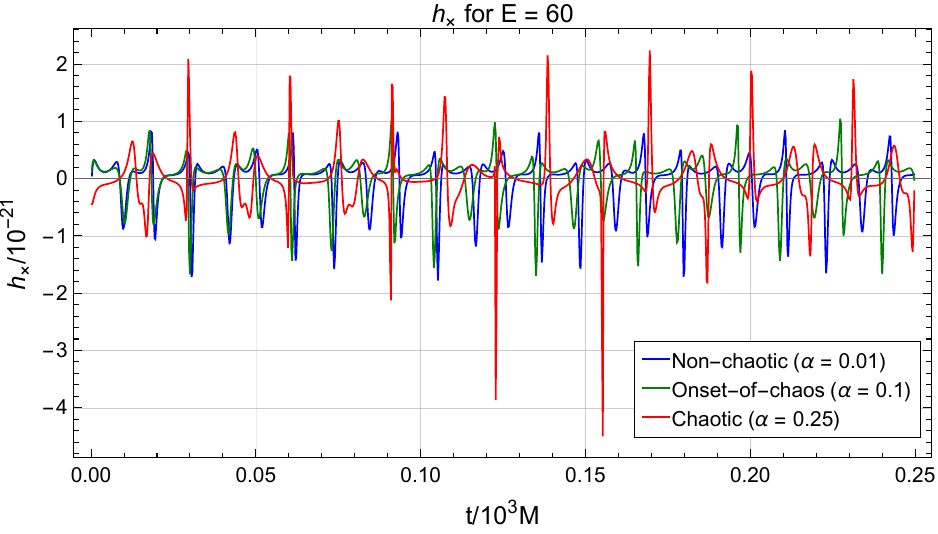}\label{Fig.2(a)}}\hfill
    \caption{The gravitational waveforms of the $+$ and $\times$ modes corresponding to the non-chaotic, onset-of-chaos and chaotic orbits for fixed $E=60$ with variations in $\alpha$. We find that the gravitational waveform from chaotic orbit and integrable orbit differ drastically.}
    \label{psm2}
\end{figure*}
\begin{figure*}[htb]
    \centering
    \subfigure[]{\includegraphics[width=0.45\textwidth]{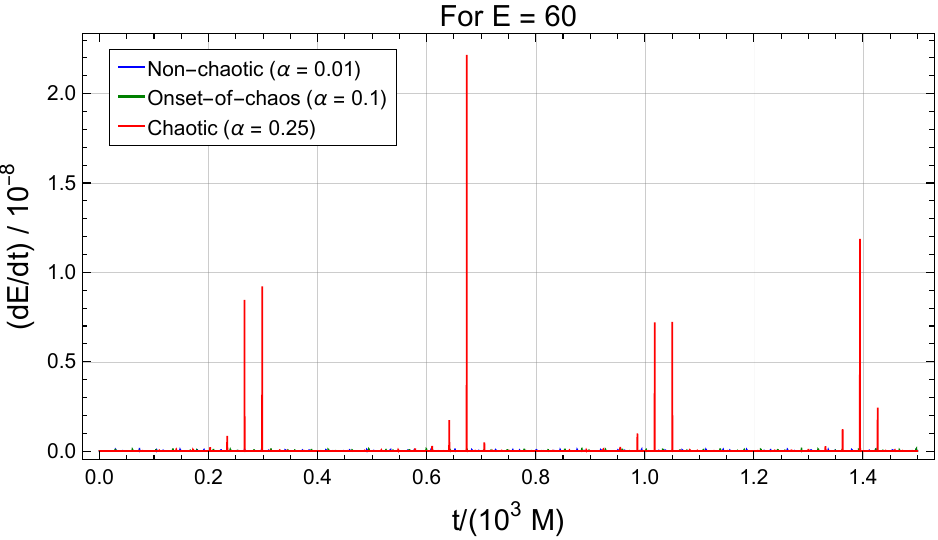}}\hfill
    \subfigure[]{\includegraphics[width=0.45\textwidth]{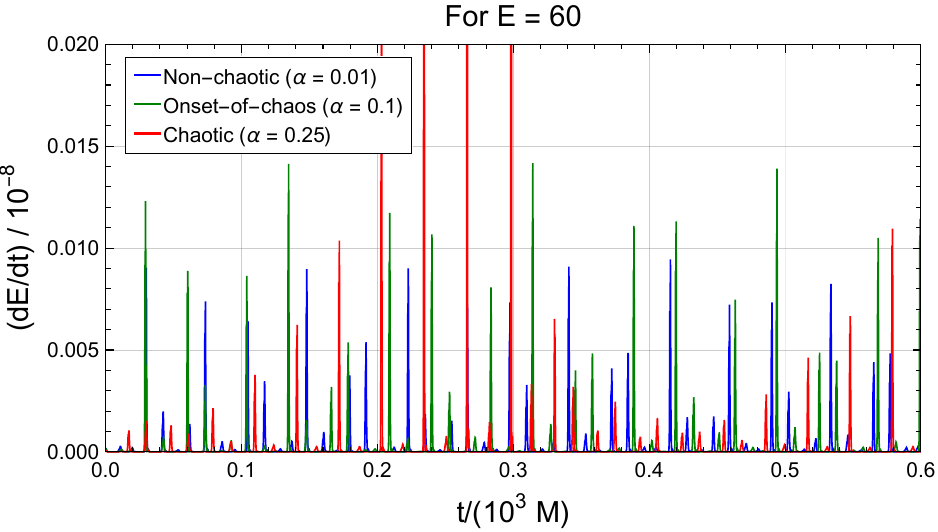}}\hfill
    \caption{The  energy emission rate of the GWs corresponding to the non-chaotic, onset-of-chaos and chaotic orbits, the right panel is the zoom-in version of the left panel. For most time in our interested time interval, the energy emission rate is relatively small thus the total energy emission is suffiencient small during the interval which guarantees the adiabatic approximation valid.}
    \label{psm3}
\end{figure*}
\subsection{\textbf{Analysis of the generated gravitational waveforms}}
To investigate the impacts of different quantum anomaly coefficients $\alpha_c$ on gravitational radiation,we consider an EMRI system composed of a small secondary compact object with mass $10M_{\odot}$ orbiting a supermassive black hole modified by quantum conformal anomalies of mass $10^6M_{\odot}$,where $M_{\odot}$ denotes the solar mass. For computational simplification,we fix the inclination angle $\iota=\pi/4$ and the longitude of periapsis within the orbital plane $\zeta=\pi/4$.  Moreover, the luminosity distance is set to $D_L$=2Gpc. 

Fig. \ref{psm2} is plotted based on the research results of black holes corrected by quantum conformal anomalies reported in our previous work \cite{An:2025xmb},and it presents gravitational waveforms for the two polarization modes ${h}_ +(f)$ and ${h}_ \times(f)$ originating from the EMRI system.These waveforms encode the gravitational influence from the black hole with conformal anomaly corrections on the formation of chaotic and non-chaotic orbits of the secondary compact object. In addition, we also compute waveforms for onset-of-chaos orbits located between non-chaotic orbits and chaotic orbits, namely the orbits where broken tori first emerge on Poincaré sections \cite{An:2025xmb}, to characterize the transition from regular ordered motion to irregular chaotic motion. We find that the amplitudes for different $\alpha$ differ drastically so the chaotic orbit near the black hole has a strong influence on the gravitational wave amplitude. As illustrated in Fig.\ref{psm2}, the overall gravitational wave amplitude remains nearly regular for non-chaotic orbits. Nevertheless, once chaos emerges in the dynamical system, the amplitudes of both polarization modes start to fluctuate,and for fully chaotic configurations, the gravitational wave amplitudes evolve in an irregular and nonstationary manner. 

When introducing the principle for generating gravitational waveforms via the numerical kludge (NK) method earlier, We state that the EMRI system we study obeys the adiabatic approximation. Specifically, we ignore the loss of energy and angular momentum of the secondary compact object as it orbits the central supermassive black hole, which arises from outward gravitational-wave emission. The angular momentum loss caused by gravitational radiation is negligible compared to the angular momentum variation induced by the harmonic oscillator potential along the 
$\phi$ direction, so if we want to focus on the leading effect, the radiation-driven angular momentum changes of the secondary body can be safely disregard. 

We then briefly explain the reason for neglecting radiative energy loss. We compute the energy emission rate for the compact object along the three distinct orbits over a finite time span. Based on the quadrupole formula derived under the weak-field regime adopted in this paper, we obtain the expression for the energy emission rate \cite{https://doi.org/10.1002/zamm.19630430611}
\begin{align}
    \frac{d\varepsilon}{dt}=&-\frac{\dddot{I}^2_{ij}}{45}
\end{align}
The negative sign in the above expression indicates energy loss via gravitational radiation. As shown in Fig.\ref{psm3},we plot the energy emission rate with fixed orbital energy $E$ and three distinct values of $\alpha$ corresponding to the three orbit types. As can be seen in Fig.\ref{psm3}, for most time in the time interval, the energy emission rate remains very small. Thus at least for the interested time interval, the integrated energy lost caused by the secondary compact object through gravitational-wave radiation is extremely small,and its impact on the orbital dynamics of the secondary body can be fully neglected. This verifies that our assumption of the adiabatic approximation for the system is reasonably valid.

\begin{figure*}[htbp]
    \subfigure[]{\includegraphics[width=0.45\textwidth]{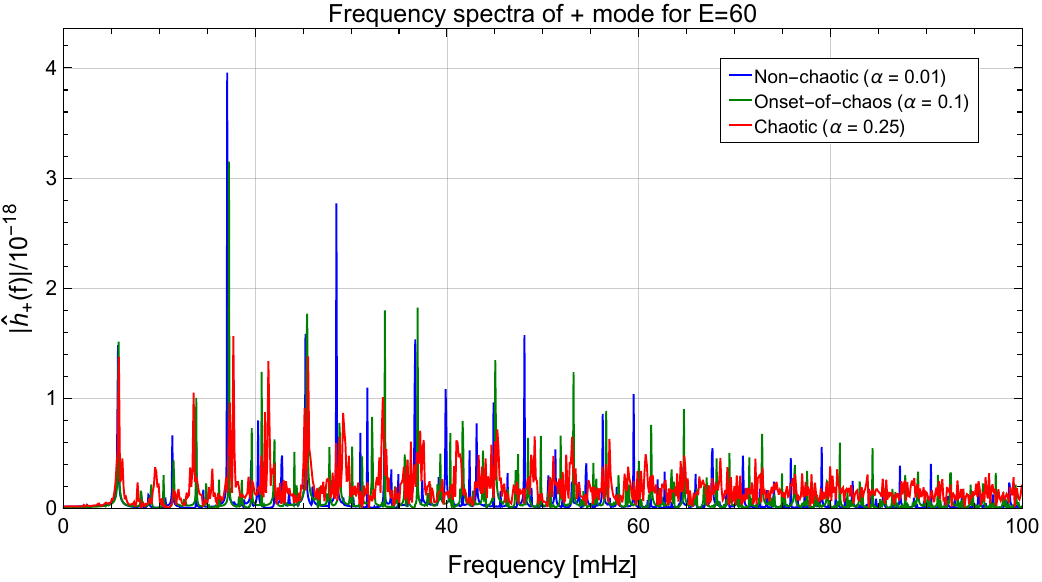}\label{Fig.4(a)}}\hfill
    \subfigure[]{\includegraphics[width=0.45\textwidth]{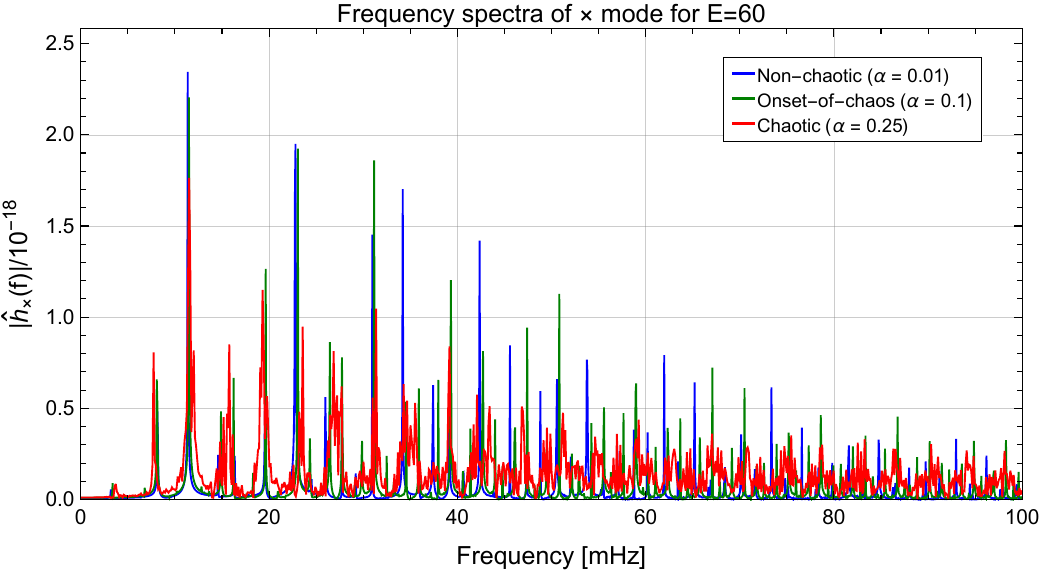}\label{Fig.4(b)}}\hfill
    \caption{The frequency spectra of different waveforms corresponding to nonchaotic, onset-of-chaos and chaotic orbitsorbits for different values of the $\alpha$. For chaotic orbit, there appears many small irregular peaks which is absent in integrable case. Thus the frequency distributaion domain is broader for chaotic orbit than integrable case.}
    \label{psm4}
\end{figure*}
\begin{figure*}[ht]
     \centering
    \subfigure[]{\includegraphics[width=0.68\textwidth]{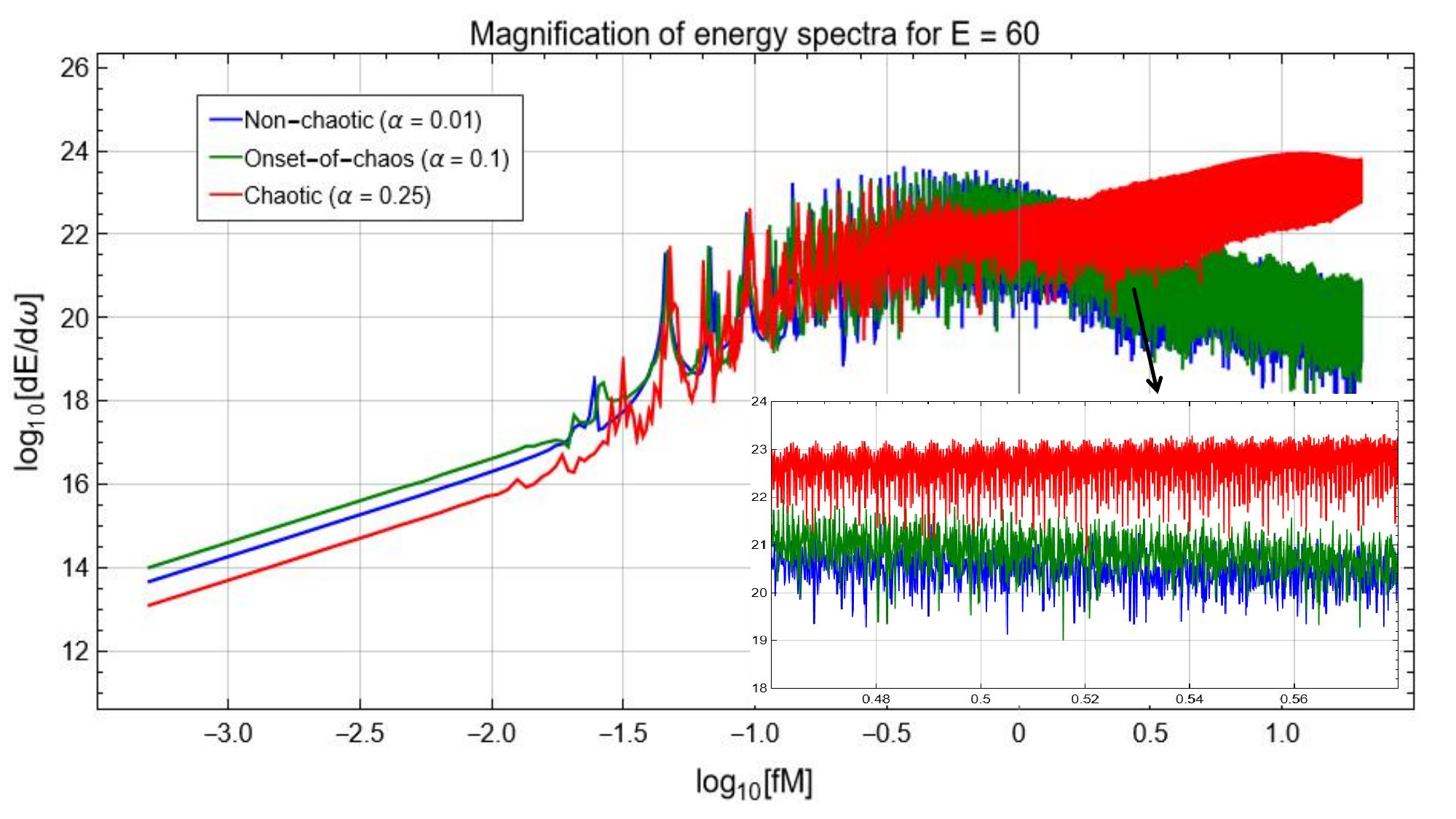}\label{Fig.5(a)}}\hfill
    \subfigure[$\alpha$ = 0.01]{\includegraphics[width=0.33\textwidth]{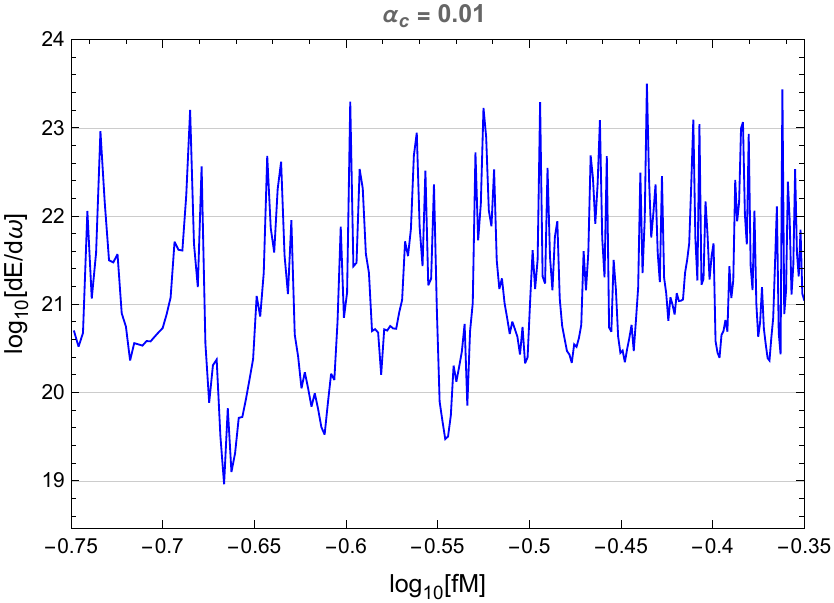}\label{Fig.5(b)}}\hfill
    \subfigure[$\alpha$ = 0.1]{\includegraphics[width=0.33\textwidth]{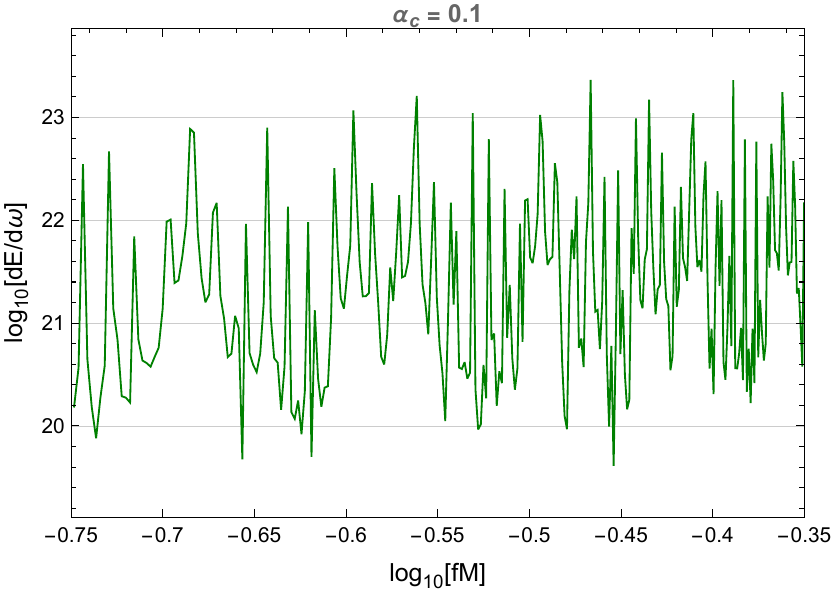}\label{Fig.5(c)}}\hfill
    \subfigure[$\alpha$ = 0.25]{\includegraphics[width=0.33\textwidth]{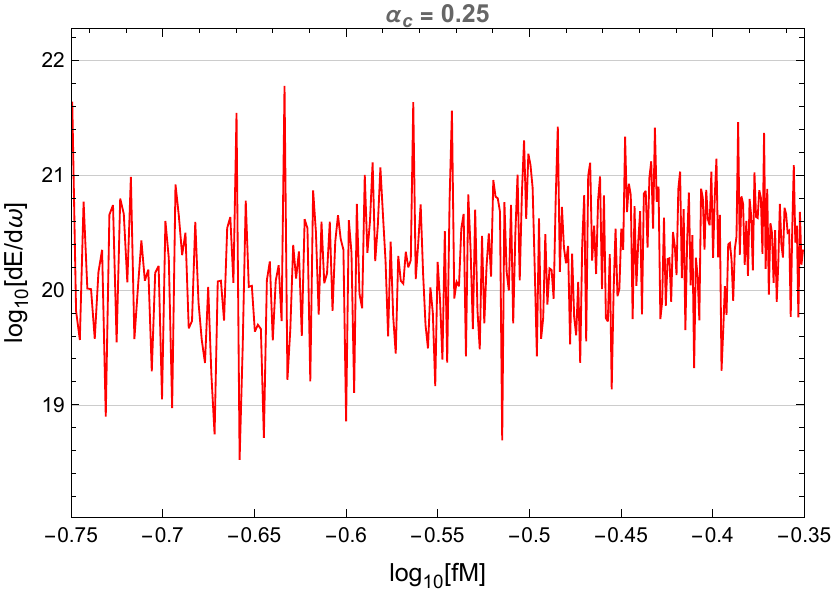}\label{Fig.5(d)}}\hfill
    \caption{Fig.\ref{Fig.5(a)} shows the GWs energy spectra corresponding to various orbits with fixed orbital energy $E$ and different values of 
$\alpha$, with an independent zoomed subplot of the region above $fM\sim10^0$ placed in the bottom right corner. Fig.\ref{Fig.5(b)}, Fig.\ref{Fig.5(c)} and Fig.\ref{Fig.5(d)} separately present the spectral segments below $fM\sim10^0$ for different $\alpha$ .
}
    \label{psm5}
\end{figure*}

\begin{figure*}[ht]
    \centering
{\includegraphics[width=0.450\textwidth]{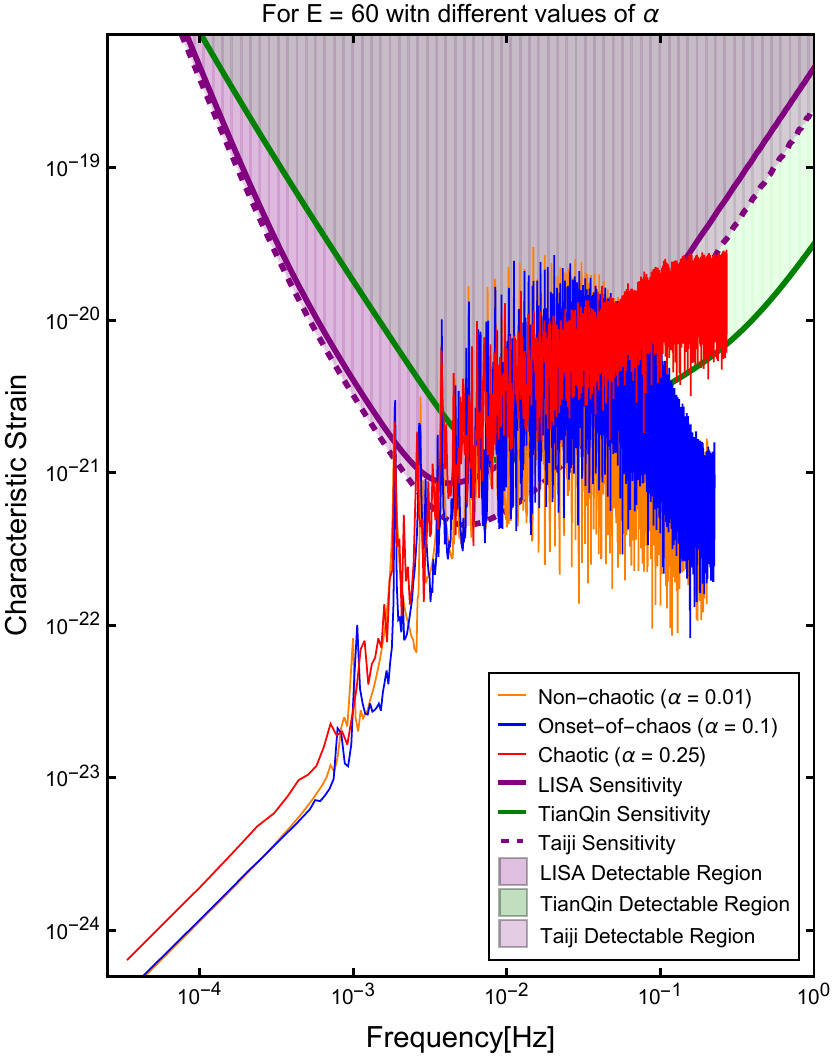}}\hfill

    \caption{The characteristic strain of GWs compared with the sensitivity curves of space-based detectors, such as LISA, Taiji, and TianQin for different orbits at the fixed $\alpha$ or at the fixed $E$. It can be seen that the signals fall within the detectable region of both LISA,Taiji and TianQin. }
    \label{psm6}
\end{figure*}
\subsection{\textbf{Analysis of the spectrum of gravitational waves}}
To further investigate the impacts of distinct quantum conformal anomaly coefficients $\alpha$ on gravitational radiation, we analyze the frequency spectra of gravitational waveforms corresponding to various orbital types based on the preceding numerical results of orbital dynamics. We first implement the discrete Fourier transform on the time-domain gravitational waveforms presented in Fig. \ref{psm2}, and the resulting frequency-domain spectra are displayed in Fig. \ref{psm4}. This figure illustrates the absolute values of  $\hat{h}_ +(f)$ and $\hat{h}_ \times(f)$, which correspond to the Fourier-transformed results of the plus and cross polarization modes, respectively.  In addition, our previous work has confirmed that the dynamical state of the system depends not only on the anomaly coefficient but also on the particle orbital energy. Since the present study mainly focuses on the impacts of different quantum anomaly coefficients on gravitational radiation, the analysis regarding the influence of varying particle energies on the system’s motion is presented in the appendix.

By examining Fig. \ref{psm4}, a prominent distinction can be observed: gravitational waveforms originating from non-chaotic orbits possess discrete frequency peaks, whereas those from chaotic orbits have many additional small peaks. In contrast to the spectra of non-chaotic and onset-of-chaos orbits, fully chaotic spectra contain numerous fine spikes and evolve into quasi-continuous spectra within a finite frequency bandwidth. 
Such spectral features extracted via Fourier analysis imply that the corresponding signals lack periodic or regular structures and instead show the intrinsic signature of chaotic dynamical systems as demonstrated in Ref.\cite{1992rcd..book.....L}. 

Our analytical results agree with previous conclusions reported in Ref.\cite{Suzuki:1999si} and Ref.\cite{Kiuchi:2004bv},which investigated gravitational radiation emitted by spinning particles orbiting Schwarzschild and Kerr black holes, as well as recent findings from Ref.\cite{Das:2025eiv} concerning chaotic dynamics within EMRI systems hosted by supermassive Schwarzschild black holes surrounded by Dehnen-(1,4,5/2) dark matter halos. The novelty of our work is that the chaotic dynamics is not induced by addition matter field but by the intrinsic quantum anomaly corrections which exist ubquitously. 

To complete the discussion, we next compute the gravitational-wave energy spectra for orbits generated with different conformal anomaly coefficients, and briefly analyze how chaos modulated by quantum conformal anomalies alters these energy spectra.
As shown in Fig. \ref{psm5},  Fig. \ref{Fig.5(a)} plots the energy spectra for various orbits corresponding to fixed $E=60$ with varying $\alpha$, respectively. 
One can observe that above the characteristic frequency $fM\sim10^0$, the spectra for different $\alpha$ values diverge substantially and develop markedly distinct evolutionary trends.
The gravitational-wave energy spectra of non-chaotic orbits and onset-of-chaos orbits gradually decay, consistent with the typical behavior of regular dynamical systems \cite{Drasco:2003ky}, which generally follows either exponential decay or steep power-law decay. In sharp contrast, the energy spectra of chaotic orbits grow with increasing frequency; this identical trend has also been reported in Ref.\cite{Das:2025eiv}, where the authors explained that tiny perturbations yield dramatic orbital deviations in chaotic systems and drive the dynamics into strongly nonlinear regimes accompanied by unpredictable orbital fluctuations, ultimately leading to rising spectral profiles under chaotic motion. 
It can be noticed that to clearly distinguish the differences in energy spectra for various orbits, we plot an enlarged view of the spectral segment above the characteristic value $fM\sim10^0$ in the bottom-right corner of Fig.\ref{Fig.5(a)}. Meanwhile, Fig.\ref{Fig.5(b)}, Fig.\ref{Fig.5(c)} and Fig.\ref{Fig.5(d)} present zoomed-in spectral regions below $fM\sim10^0$ for orbits with different values of $\alpha$, respectively.
From Fig.\ref{Fig.5(b)},multiple sharp spectral peaks emerge at discrete characteristic frequencies, a signature naturally originating from the regular periodic orbital motion as previously demonstrated in Ref.\cite{Das:2025eiv}. In contrast, the spectrum in Fig.\ref{Fig.5(d)} contains numerous irregular peaks lacking pronounced sharpness; substantial superposition of dense spiky features smears the spectrum into a broad quasi-continuous profile. 
\subsection{\textbf{Scopes of detectability by future-based detection antennas}}
In the preceding sections, we first clarified the influence of conformal anomalies on chaotic dynamics and then investigated the imprints of conformal-anomaly-modulated chaotic motion embedded within gravitational-wave emission. Specifically, we systematically compared the gravitational waveforms, frequency spectra and energy spectra radiated from non-chaotic orbits, onset-of-chaos orbits and chaotic orbits in EMRI systems, confirming that all these spectral profiles are substantially altered by chaos. 

To determine whether such an effect can be directly detected by future space-based GW detectors including LISA, Taiji and TianQin, we compare the characteristic strain of gravitational waves with the sensitivity curves of these detectors to judge whether the signals induced by this effect are detectable in future space-based gravitational-wave observatories
To address the above issue, we first specify the sensitivity curves of the considered detectors, namely LISA, Taiji and TianQin, which are required for the subsequent verification analysis, and we present and briefly introduce these curves in the following. 

We adopt the noise-free sensitivity curve of LISA from Ref.\cite{Robson:2018ifk}, which neglects confusion noise and provides a well-behaved sensitivity model adequate for our analysis: \begin{equation}\label{Lisa}
    S_n^L(f)=\frac{10}{3L^2}\Big(P_{OMS}+\frac{2P_{acc}}{(2\pi f)^2}(1+\cos^2(\frac{f}{f^*}))\Big)\times \Big(1+0.6(\frac{f}{f^*})^2\Big),
\end{equation}
where $L=2.5\times10^9\mathrm{m}$ specifies the arm length for LISA, $f^*=c/2\pi L\sim 19.09 mHz$ defines the transfer frequency. Finally, $P_{OMS}$ and $P_{acc}$  denote the single-link optical metrology noise component and test mass acceleration noise, respectively, whose explicit expressions are given below \cite{Robson:2018ifk}: \begin{align}
P_{\mathrm{OMS}}(f)=&\left(1.5\times10^{-11}\mathrm{m}\right)^2\times\left(1+\left(\frac{2\mathrm{mHz}}{f}\right)^4\right)\mathrm{Hz}^{-1},\\
P_{\mathrm{acc}}(f)=&\left(3\times10^{-15}\mathrm{ms^{-2}}\right)^2\left(1+\left(\frac{0.4\mathrm{mHz}}{f}\right)^2\right)\notag\\ 
&\times \left(1+\left(\frac{f}{\mathrm{8mHz}}\right)^4\right)\mathrm{Hz}^{-1},
\end{align}
The TianQin sensitivity curve follows the formulation as \cite{Li:2024rnk,TianQin:2020hid}
\begin{equation}\label{TianQin}
    S_N(f)=\frac{10}{3L^2}\Big(S_x+\frac{4S_a}{(2\pi f)^4}(1+\frac{10^{-4}\mathrm{Hz}}{f}\Big)\times \Big(1+0.6(\frac{f}{f_*})^2\Big)
\end{equation}
where $L=1.7\times10^8\mathrm{m}$ specifies the arm length of TianQin,$S_a=1\times10^{-30}\mathrm{m^2s^{-4} Hz^{-1}}$ quantifies the acceleration noise,$S_x=1\times10^{-24}\mathrm{m^2 Hz^{-1}}$ describes the displacement measurement noise,$f_*=c/2\pi L\sim 280.66 mHz$ is the transfer frequency of TianQin.

Lastly, we adopt the sensitivity curve of Taiji from the model presented in Ref.\cite{Hu:2017mde,Liu:2023qap}, which shares an identical fundamental structure with the formula in \label{Lisa}, as expressed below \cite{Liu:2023qap}: \begin{align}\label{Taiji}
    S_n^T(f)=\frac{10}{3L^2}\Big(P_{dp}+\frac{2P_{acc}}{(2\pi f)^4}(1+\cos^2(\frac{f}{f^*}))\Big)\times \Big(1+0.6(\frac{f}{f^*})^2\Big),
\end{align}
where the arm length is $L=3\times 10^9\mathrm{m}$  for Taiji,the transfer frequency is $f_*=c/2\pi L\sim 15.90 mHz$. Here,$P_{dp}$ and $P_{acc}$ represent the power spectral density of the displacement noise and the acceleration noise, respectively, having the following forms as \cite{Liu:2023qap}\begin{align}
P_{\mathrm{dp}}(f)=&\left(8\times10^{-12}\mathrm{m}\right)^2\times\left(1+\left(\frac{2\mathrm{mHz}}{f}\right)^4\right)\mathrm{Hz}^{-1},\\
P_{\mathrm{acc}}(f)=&\left(3\times10^{-15}\mathrm{ms^{-2}}\right)^2\left(1+\left(\frac{0.4\mathrm{mHz}}{f}\right)^2\right)\notag\\
&\times \left(1+\left(\frac{f}{\mathrm{8mHz}}\right)^4\right)\mathrm{Hz}^{-1},
\end{align}

With the sensitivity curves of future space-based GW detectors available, we then compute the corresponding characteristic strain using the gravitational-wave spectra emitted from various orbits around quantum-conformal-anomaly-corrected black holes derived in previous sections. The definition of characteristic strain is given as follows:\begin{equation}
    h_c(f)=2|f|\sqrt{|\hat{h}_+(f)|^2+|\hat{h}_\times(f)|^2}
\end{equation} 

With this definition, we proceed to calculate the characteristic strain of gravitational waves radiated from three distinct classes of orbits (non-chaotic orbits, onset-of-chaos orbits and chaotic orbits), where each orbit corresponds to a specific set of conformal anomaly coefficient $\alpha$ and orbital energy $E$. To examine whether gravitational-wave signals originating from conformal-anomaly-driven chaotic motion are detectable by future space-based GW detectors, we overlay the calculated characteristic strain and the sensitivity curves of the three detectors in a single figure for direct comparison. Notably, the detector sensitivity curves must be nondimensionalized into the forms $\sqrt{fS_n^L}$ ,$\sqrt{fS_N}$ and $\sqrt{fS_n^T}$ before comparison against our computed characteristic strain, and the resulting comparison is displayed in Fig.\ref{psm6}. 

It can be observed that portions of these strain curves rise above the sensitivity thresholds of LISA, Taiji and TianQin, demonstrating that gravitational-wave signals from EMRI systems surrounding quantum-conformal-anomaly-corrected black holes are promising targets for detection by forthcoming space-based GW observatories. In particular, signals sourced from chaotic orbits carry imprints of conformal anomalies on chaotic dynamics; accordingly, observational measurements of such gravitational waves enable indirect detection and constraint of conformal anomaly effects. 

Furthermore, prominent discrepancies emerge at frequencies well around $10^{-1} \mathrm{Hz}$, the characteristic strain from non-chaotic orbits and orbits at the chaos onset declines sharply at high frequencies, whereas fully developed chaotic orbits show no such rapid drop. This behavior replicates the spectral trends identified in our earlier energy-spectrum analysis and originates from the strong nonlinearity inherent to fully chaotic dynamical systems. This phenomenon is expected to be found in future observation such as TianQin detector.

\section{Conclusion and Discussion:}\label{sec4}
In this work, we investigate gravitational waves emitted from EMRI systems consisting of supermassive black holes with quantum conformal anomaly corrections and a small surrounding steller-mass objects. Built upon our prior research on the chaotic orbital dynamics of test particles in the background of quantum-conformal-anomaly-corrected black holes, 
we further found that such an integrable-chaotic dynamical transition would leave observable imprints on the resulting gravitational radiation. We adopt the numerical kludge approach to construct the gravitational-wave signals emitted by these systems. Comparative analysis verifies that the dynamical transition is encoded within waveforms,frequency distribution and energy spectra of the emitted gravitational waves. Our numerical results demonstrate that gravitational waves from conformal-anomaly-induced chaotic orbits feature irregular, strongly time-varying amplitudes, along with abundant fine spectral spikes and extended continuous spectral profiles in both frequency and energy domains. Moreover, after computing characteristic strain from the obtained frequency spectra, we perform comparisons against the sensitivity curves of future space-based GW detectors including LISA, Taiji and TianQin. The comparison confirms that these planned detectors possess sufficient sensitivity to capture gravitational-wave signals originating from conformal-anomaly-affected chaotic EMRI configurations, which establishes a direct observational pathway for probing conformal anomaly effects in astrophysical environments in future experiments. 

There are also many issues derserving to be persued in the future. Firstly, our gravitational-wave computations are performed under the adiabatic approximation with a limited observation time duration, neglecting the backreaction of gravitational radiation on orbital evolution. Therefore, future investigations should address the combined impacts of conformal anomalies and gravitational-wave backreaction on long-term chaotic orbital evolution. Additionally, our current calculation only accounts for quadrupole gravitational-wave contributions while neglecting higher multipole corrections, inevitably introducing numerical inaccuracies. Follow-up studies can incorporate higher-order multipole terms or adopt more rigorous, comprehensive computational frameworks to improve the existing results.
These issues remain to be explored in future research.

\section*{Acknowledgements}
This work is supported by the Natural Science Foundation of Jiangsu Province under Grant No. BK20241376 and the National Natural Science Foundation of China (NSFC) under Grants No.12405066. 
\appendix
\section{Gravitational waves for various energies:}
The main body above primarily discusses how gravitational waveforms are affected by the conformal anomaly coefficient $\alpha$. To complement the main content, we supplement the analysis of how the compact object’s orbital energy $E$ impacts gravitational waveforms in the appendix. Following the same analytical approach adopted earlier, we plot the corresponding figures as presented below.

First, we plot time-domain diagrams for orbits with different energies in Fig.\ref{Fig.A1}. Similarly, the overall amplitude of gravitational waves from non-chaotic orbits at 
$E=50$ remains nearly constant and shows periodic features. When 
$E=70$, the system enters the chaos onset regime, and the amplitude begins to fluctuate. For fully chaotic motion at $E=80$, the gravitational-wave amplitude displays irregular, time-varying variations with no periodic structure. We then perform Fourier transforms to obtain frequency-domain spectra, as illustrated in Fig.\ref{Fig.A3}. As the orbit transitions from non-chaotic to chaotic, the frequency-domain profile evolves from discrete characteristic frequencies into a finite-band quasi-continuous spectrum filled with numerous fine spikes.

The energy spectra plotted in Fig. \ref{Fig.A4} show that orbits transition from regular to chaotic as the orbital energy increases. This transition is manifested as elevated curves and a large number of overlapping irregular fine spikes in the enlarged subplots, forming a broad quasi-continuous spectrum. Finally, we conduct a simple detectability assessment by overlaying the sensitivity curves of the three detectors with the characteristic strain curves of these gravitational waves, as shown in Fig. \ref{Fig.A5}. It is clear that portions of these signals lie above the detector sensitivity curves, which implies that signals from such systems will be detectable in future observations.
\setcounter{figure}{0}
\renewcommand{\thefigure}{A\arabic{figure}}

\begin{figure*}[ht]
    \centering
    \subfigure[]{\includegraphics[width=0.450\textwidth]{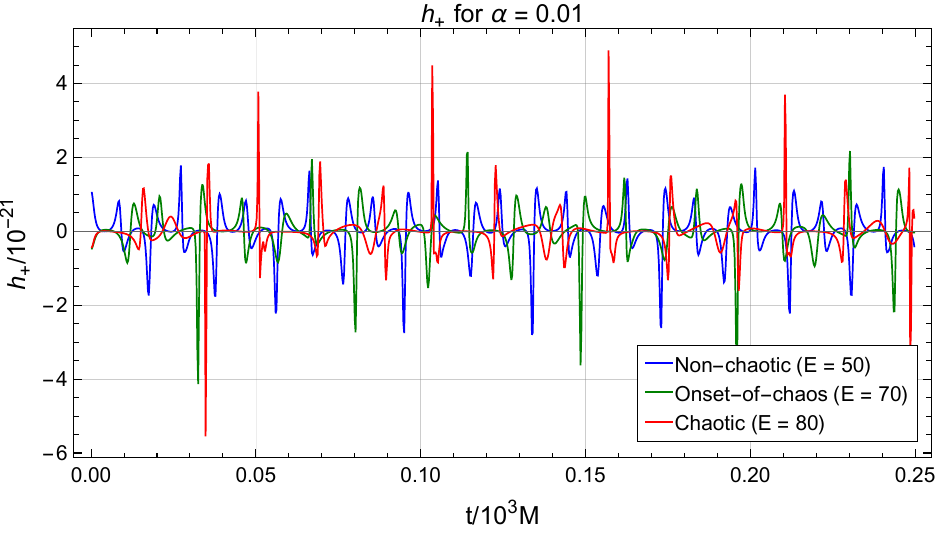}}\hfill
    \subfigure[]{\includegraphics[width=0.45\textwidth]{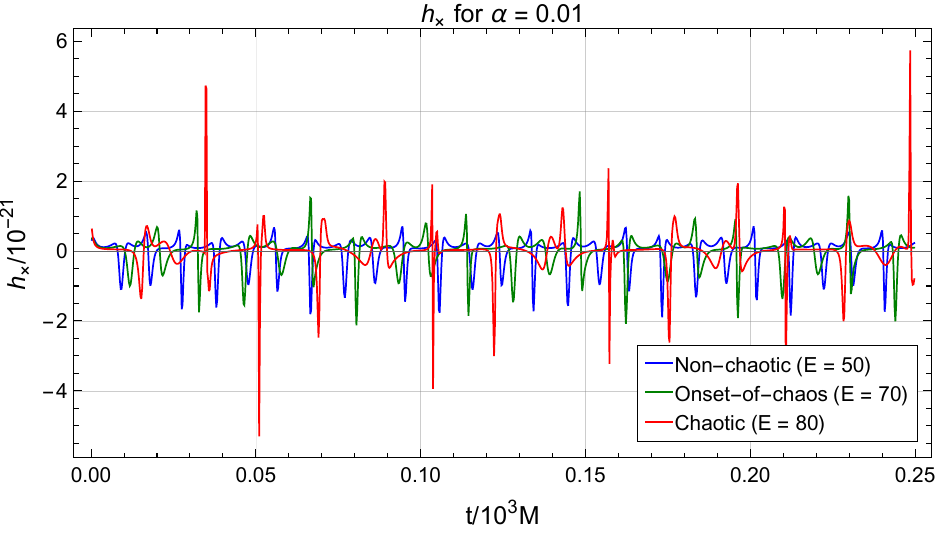}}\hfill
     \caption{The gravitational waveforms of the $+$ and $\times$ modes corresponding to the non-chaotic, onset-of-chaos and chaotic orbits for fixed $\alpha = 0.01$ with variations in $E$.}
    \label{Fig.A1}
\end{figure*}


\begin{figure*}[ht]
    \subfigure[]{\includegraphics[width=0.45\textwidth]{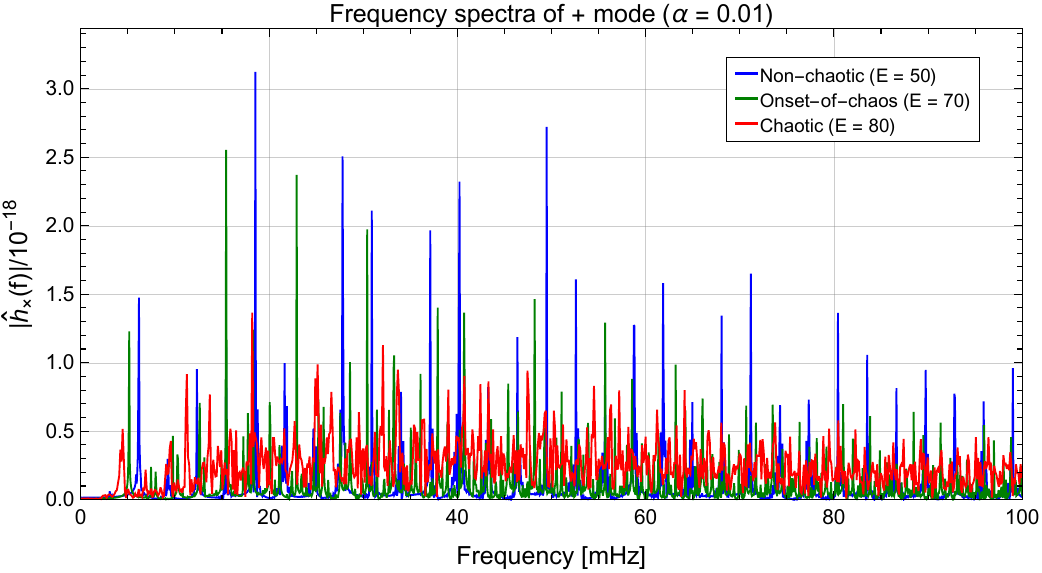}\label{Fig.A3(a)}}\hfill
    \subfigure[]{\includegraphics[width=0.45\textwidth]{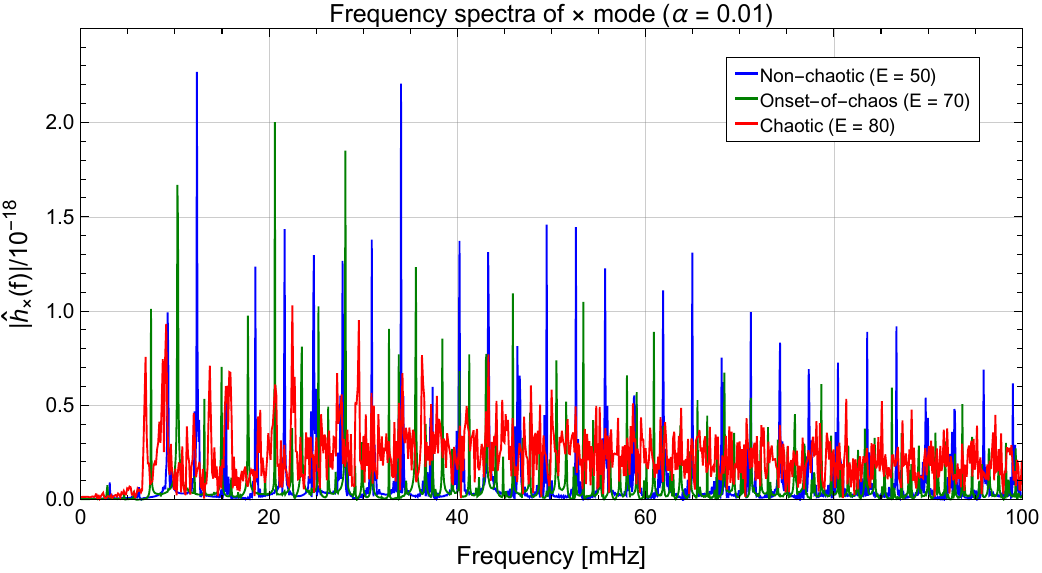}\label{Fig.A3(b)}}\hfill
    \caption{The frequency spectra of different waveforms corresponding to nonchaotic, onset-of-chaos and chaotic orbitsorbits for
different values of the $E$. }
    \label{Fig.A3}
\end{figure*}
\begin{figure*}[htbp]
     \centering
    \subfigure[]{\includegraphics[width=0.68\textwidth]{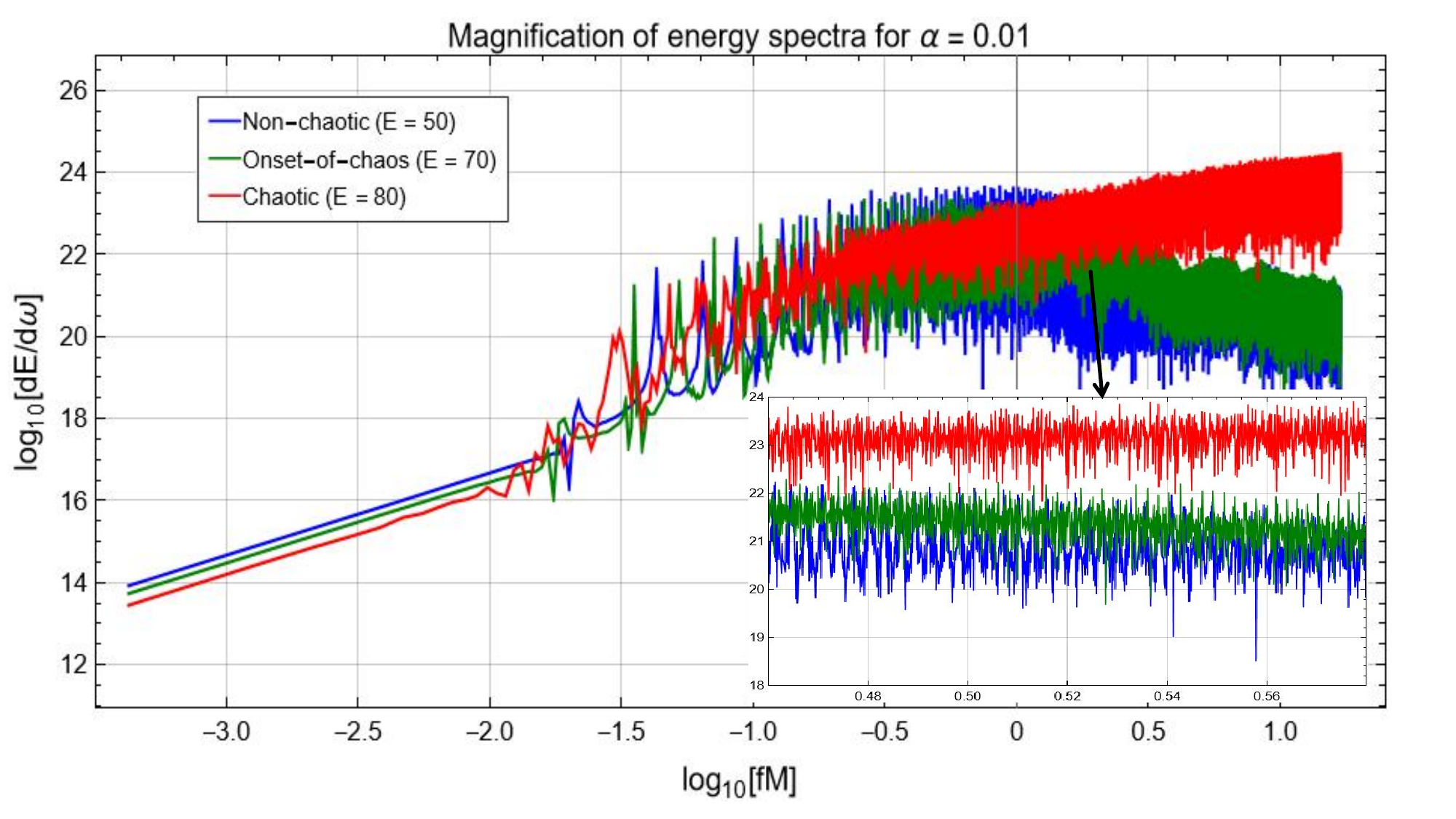}\label{Fig.A4(a)}}\hfill
    \subfigure[$E=50$]{\includegraphics[width=0.33\textwidth]{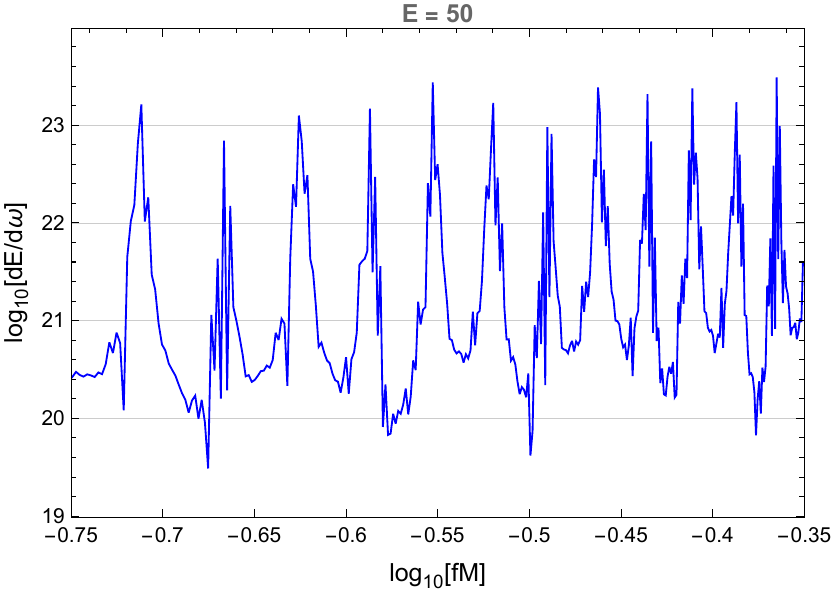}\label{Fig.A4(b)}}\hfill
    \subfigure[$E=70$ ]{\includegraphics[width=0.33\textwidth]{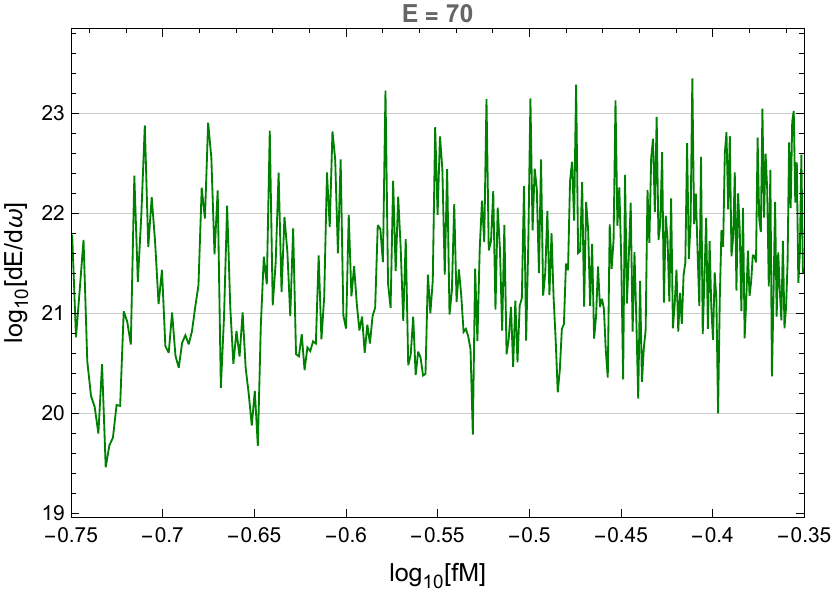}\label{Fig.A4(c)}}\hfill
    \subfigure[$E=80$]{\includegraphics[width=0.33\textwidth]{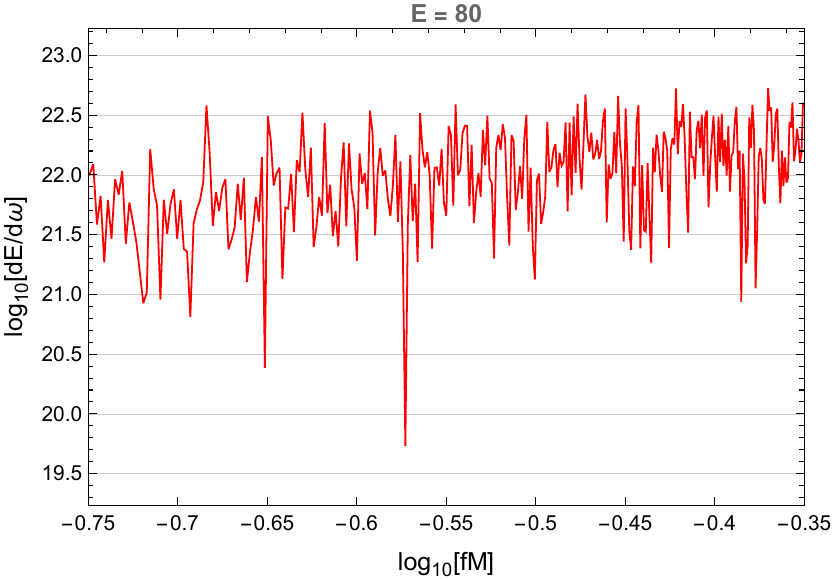}\label{Fig.A4(d)}}\hfill
    \caption{Fig.\ref{Fig.A4(a)} shows the GWs energy spectra corresponding to various orbits with fixed orbital  $\alpha$ and different values of 
energy $E$, with an independent zoomed subplot of the region above $fM\sim10^0$ placed in the bottom right corner. Fig.\ref{Fig.A4(b)}, Fig.\ref{Fig.A4(c)} and Fig.\ref{Fig.A4(d)} separately present the spectral segments below $fM\sim10^0$ for different $\alpha$.}
    \label{Fig.A4}
\end{figure*}

\begin{figure*}[ht]
    \centering
{\includegraphics[width=0.48\textwidth]{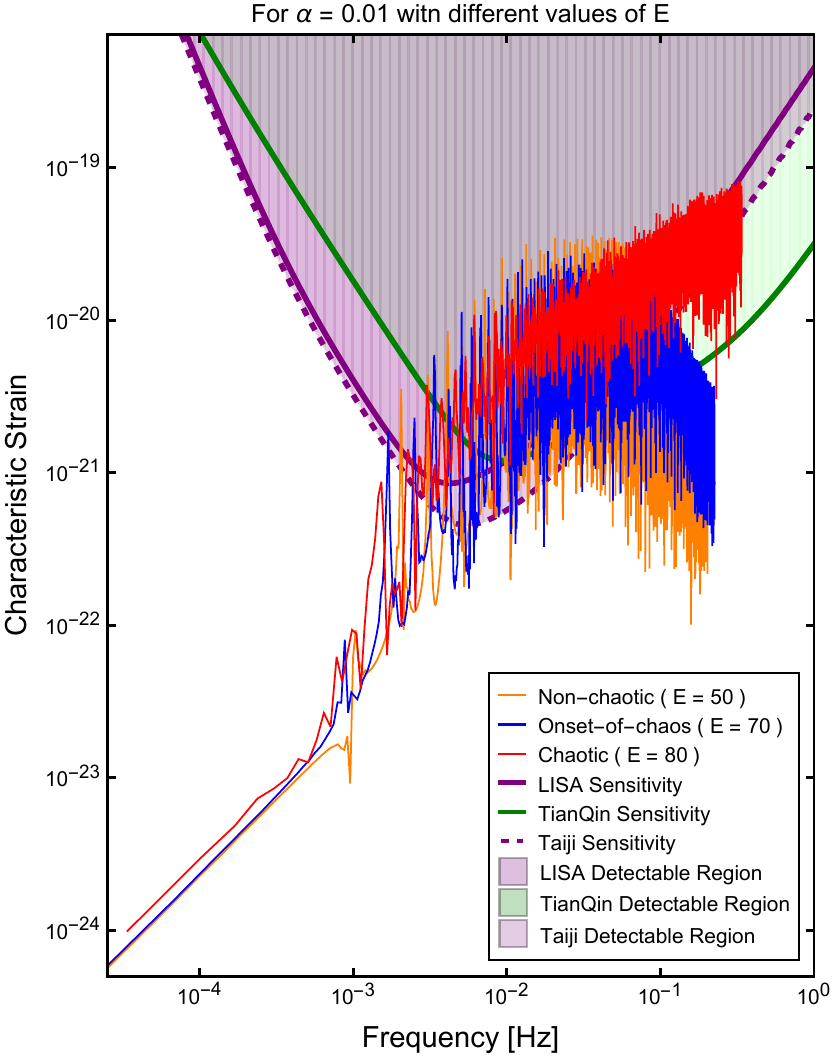}}\hfill
    \caption{The characteristic strain of GWs compared with the sensitivity curves of space-based detectors, such as LISA, Taiji, and TianQin for different orbits at the fixed $\alpha$ or at the fixed $E$}
    \label{Fig.A5}
\end{figure*}

\bibliographystyle{elsarticle-num} 
\biboptions{sort&compress}
\bibliography{msc}

@article{An:2025xmb,
    author = "An, Yu-Sen and Zhang, Wei-Hao",
    title = "{Probing quantum anomaly corrections on black hole physics through chaos}",
    eprint = "2512.19163",
    archivePrefix = "arXiv",
    primaryClass = "gr-qc",
    month = "12",
    year = "2025"
}

@article{Peters:1964zz,
    author = "Peters, P. C.",
    title = "{Gravitational Radiation and the Motion of Two Point Masses}",
    doi = "10.1103/PhysRev.136.B1224",
    journal = "Phys. Rev.",
    volume = "136",
    pages = "B1224--B1232",
    year = "1964"
}

@article{Einstein:1937qu,
    author = "Einstein, Albert and Rosen, N.",
    title = "{On Gravitational waves}",
    doi = "10.1016/S0016-0032(37)90583-0",
    journal = "J. Franklin Inst.",
    volume = "223",
    pages = "43--54",
    year = "1937"
}

@article{LIGOScientific:2016aoc,
    author = "Abbott, B. P. and others",
    collaboration = "LIGO Scientific, Virgo",
    title = "{Observation of Gravitational Waves from a Binary Black Hole Merger}",
    eprint = "1602.03837",
    archivePrefix = "arXiv",
    primaryClass = "gr-qc",
    reportNumber = "LIGO-P150914",
    doi = "10.1103/PhysRevLett.116.061102",
    journal = "Phys. Rev. Lett.",
    volume = "116",
    number = "6",
    pages = "061102",
    year = "2016"
}

@article{LIGOScientific:2019zcs,
    author = "Abbott, B. P. and others",
    collaboration = "LIGO Scientific, Virgo, VIRGO",
    title = "{A Gravitational-wave Measurement of the Hubble Constant Following the Second Observing Run of Advanced LIGO and Virgo}",
    eprint = "1908.06060",
    archivePrefix = "arXiv",
    primaryClass = "astro-ph.CO",
    reportNumber = "LIGO-P1900015",
    doi = "10.3847/1538-4357/abdcb7",
    journal = "Astrophys. J.",
    volume = "909",
    number = "2",
    pages = "218",
    year = "2021"
}

@article{LIGOScientific:2016emj,
    author = "Abbott, B. P. and others",
    collaboration = "LIGO Scientific, Virgo",
    title = "{GW150914: The Advanced LIGO Detectors in the Era of First Discoveries}",
    eprint = "1602.03838",
    archivePrefix = "arXiv",
    primaryClass = "gr-qc",
    reportNumber = "LIGO-P1500237",
    doi = "10.1103/PhysRevLett.116.131103",
    journal = "Phys. Rev. Lett.",
    volume = "116",
    number = "13",
    pages = "131103",
    year = "2016"
}

@article{EventHorizonTelescope:2019uob,
    author = "Akiyama, Kazunori and others",
    collaboration = "Event Horizon Telescope",
    title = "{First M87 Event Horizon Telescope Results. II. Array and Instrumentation}",
    eprint = "1906.11239",
    archivePrefix = "arXiv",
    primaryClass = "astro-ph.IM",
    doi = "10.3847/2041-8213/ab0c96",
    journal = "Astrophys. J. Lett.",
    volume = "875",
    number = "1",
    pages = "L2",
    year = "2019"
}

@article{EventHorizonTelescope:2022wkp,
    author = "Akiyama, Kazunori and others",
    collaboration = "Event Horizon Telescope",
    title = "{First Sagittarius A* Event Horizon Telescope Results. I. The Shadow of the Supermassive Black Hole in the Center of the Milky Way}",
    eprint = "2311.08680",
    archivePrefix = "arXiv",
    primaryClass = "astro-ph.HE",
    doi = "10.3847/2041-8213/ac6674",
    journal = "Astrophys. J. Lett.",
    volume = "930",
    number = "2",
    pages = "L12",
    year = "2022"
}

@article{EventHorizonTelescope:2022apq,
    author = "Akiyama, Kazunori and others",
    collaboration = "Event Horizon Telescope",
    title = "{First Sagittarius A* Event Horizon Telescope Results. II. EHT and Multiwavelength Observations, Data Processing, and Calibration}",
    eprint = "2311.08679",
    archivePrefix = "arXiv",
    primaryClass = "astro-ph.HE",
    reportNumber = "FERMILAB-PUB-22-418-PPD",
    doi = "10.3847/2041-8213/ac6675",
    journal = "Astrophys. J. Lett.",
    volume = "930",
    number = "2",
    pages = "L13",
    year = "2022"
}

@article{Ozel:2010bz,
    author = "Ozel, Feryal and Psaltis, Dimitrios and Ransom, Scott and Demorest, Paul and Alford, Mark",
    title = "{The Massive Pulsar PSR J1614-2230: Linking Quantum Chromodynamics, Gamma-ray Bursts, and Gravitational Wave Astronomy}",
    eprint = "1010.5790",
    archivePrefix = "arXiv",
    primaryClass = "astro-ph.HE",
    doi = "10.1088/2041-8205/724/2/L199",
    journal = "Astrophys. J. Lett.",
    volume = "724",
    pages = "L199--L202",
    year = "2010"
}

@article{Hashimoto:2016dfz,
    author = "Hashimoto, Koji and Tanahashi, Norihiro",
    title = "{Universality in Chaos of Particle Motion near Black Hole Horizon}",
    eprint = "1610.06070",
    archivePrefix = "arXiv",
    primaryClass = "hep-th",
    reportNumber = "OU-HET-911",
    doi = "10.1103/PhysRevD.95.024007",
    journal = "Phys. Rev. D",
    volume = "95",
    number = "2",
    pages = "024007",
    year = "2017"
}

@article{Dalui:2018qqv,
    author = "Dalui, Surojit and Majhi, Bibhas Ranjan and Mishra, Pankaj",
    title = "{Presence of horizon makes particle motion chaotic}",
    eprint = "1803.06527",
    archivePrefix = "arXiv",
    primaryClass = "gr-qc",
    doi = "10.1016/j.physletb.2018.11.050",
    journal = "Phys. Lett. B",
    volume = "788",
    pages = "486--493",
    year = "2019"
}

@article{Bera:2021lgw,
    author = "Bera, Avijit and Dalui, Surojit and Ghosh, Subir and Vagenas, Elias C.",
    title = "{Quantum corrections enhance chaos: Study of particle motion near a generalized Schwarzschild black hole}",
    eprint = "2109.00330",
    archivePrefix = "arXiv",
    primaryClass = "gr-qc",
    doi = "10.1016/j.physletb.2022.137033",
    journal = "Phys. Lett. B",
    volume = "829",
    pages = "137033",
    year = "2022"
}

@article{Maldacena:2015waa,
    author = "Maldacena, Juan and Shenker, Stephen H. and Stanford, Douglas",
    title = "{A bound on chaos}",
    eprint = "1503.01409",
    archivePrefix = "arXiv",
    primaryClass = "hep-th",
    doi = "10.1007/JHEP08(2016)106",
    journal = "JHEP",
    volume = "08",
    pages = "106",
    year = "2016"
}

@article{Yu:2023spr,
    author = "Yu, Chengye and Chen, Deyou and Mu, Benrong and He, Yucheng",
    title = "{Violating the chaos bound in five-dimensional, charged, rotating Einstein-Maxwell-Chern-Simons black holes}",
    doi = "10.1016/j.nuclphysb.2023.116093",
    journal = "Nucl. Phys. B",
    volume = "987",
    pages = "116093",
    year = "2023"
}

@article{Dutta:2024rta,
    author = "Dutta, Pinaki and Panigrahi, Kamal L. and Singh, Balbeer",
    title = "{Chaos bound and its violation in black p-brane}",
    eprint = "2408.14056",
    archivePrefix = "arXiv",
    primaryClass = "hep-th",
    doi = "10.1007/JHEP02(2025)043",
    journal = "JHEP",
    volume = "02",
    pages = "043",
    year = "2025"
}

@article{Gao:2022ybw,
    author = "Gao, Chuanhong and Chen, Deyou and Yu, Chengye and Wang, Peng",
    title = "{Chaos bound and its violation in charged Kiselev black hole}",
    eprint = "2204.07983",
    archivePrefix = "arXiv",
    primaryClass = "gr-qc",
    doi = "10.1016/j.physletb.2022.137343",
    journal = "Phys. Lett. B",
    volume = "833",
    pages = "137343",
    year = "2022"
}

@article{Gwak:2022xje,
    author = "Gwak, Bogeun and Kan, Naoto and Lee, Bum-Hoon and Lee, Hocheol",
    title = "{Violation of bound on chaos for charged probe in Kerr-Newman-AdS black hole}",
    eprint = "2203.07298",
    archivePrefix = "arXiv",
    primaryClass = "gr-qc",
    doi = "10.1007/JHEP09(2022)026",
    journal = "JHEP",
    volume = "09",
    pages = "026",
    year = "2022"
}

@article{Lei:2020clg,
    author = "Lei, Yu-Qi and Ge, Xian-Hui and Ran, Cheng",
    title = "{Chaos of particle motion near a black hole with quasitopological electromagnetism}",
    eprint = "2008.01384",
    archivePrefix = "arXiv",
    primaryClass = "hep-th",
    doi = "10.1103/PhysRevD.104.046020",
    journal = "Phys. Rev. D",
    volume = "104",
    number = "4",
    pages = "046020",
    year = "2021"
}

@article{Xie:2025auj,
    author = "Xie, Hao and Yang, Si-Jiang",
    title = "{Probing phase transitions of regular black holes in anti-de Sitter space with Lyapunov exponent}",
    eprint = "2510.23387",
    archivePrefix = "arXiv",
    primaryClass = "gr-qc",
    doi = "10.1140/epjc/s10052-025-15111-y",
    journal = "Eur. Phys. J. C",
    volume = "85",
    number = "12",
    pages = "1374",
    year = "2025"
}

@article{Zhang:2025cdx,
    author = "Zhang, Shi-Hao and Zhao, Zi-Qiang and Li, Zi-Yuan and Zhang, Jing-Fei and Zhang, Xin",
    title = "{Gaussian curvature and Lyapunov exponent as probes of black hole phase transitions}",
    eprint = "2509.05103",
    archivePrefix = "arXiv",
    primaryClass = "gr-qc",
    month = "9",
    year = "2025"
}

@article{Shukla:2024tkw,
    author = "Shukla, Bhaskar and Das, Pranaya Pratik and Dudal, David and Mahapatra, Subhash",
    title = "{Interplay between the Lyapunov exponents and phase transitions of charged AdS black holes}",
    eprint = "2404.02095",
    archivePrefix = "arXiv",
    primaryClass = "hep-th",
    doi = "10.1103/PhysRevD.110.024068",
    journal = "Phys. Rev. D",
    volume = "110",
    number = "2",
    pages = "024068",
    year = "2024"
}

@article{Guo:2022kio,
    author = "Guo, Xiaobo and Lu, Yuhang and Mu, Benrong and Wang, Peng",
    title = "{Probing phase structure of black holes with Lyapunov exponents}",
    eprint = "2205.02122",
    archivePrefix = "arXiv",
    primaryClass = "gr-qc",
    reportNumber = "CTP-SCU/2022008",
    doi = "10.1007/JHEP08(2022)153",
    journal = "JHEP",
    volume = "08",
    pages = "153",
    year = "2022"
}

@article{Das:2024iuf,
    author = "Das, Surajit and Dalui, Surojit and Samanta, Rickmoy",
    title = "{Near-horizon chaos beyond Einstein gravity}",
    eprint = "2405.09945",
    archivePrefix = "arXiv",
    primaryClass = "gr-qc",
    doi = "10.1103/PhysRevD.110.124037",
    journal = "Phys. Rev. D",
    volume = "110",
    number = "12",
    pages = "124037",
    year = "2024"
}

@article{Shenker:2013pqa,
    author = "Shenker, Stephen H. and Stanford, Douglas",
    title = "{Black holes and the butterfly effect}",
    eprint = "1306.0622",
    archivePrefix = "arXiv",
    primaryClass = "hep-th",
    reportNumber = "SU-ITP-13-08",
    doi = "10.1007/JHEP03(2014)067",
    journal = "JHEP",
    volume = "03",
    pages = "067",
    year = "2014"
}

@article{Roberts:2014isa,
    author = "Roberts, Daniel A. and Stanford, Douglas and Susskind, Leonard",
    title = "{Localized shocks}",
    eprint = "1409.8180",
    archivePrefix = "arXiv",
    primaryClass = "hep-th",
    reportNumber = "MIT-CTP-4594, SU-ITP-14-20",
    doi = "10.1007/JHEP03(2015)051",
    journal = "JHEP",
    volume = "03",
    pages = "051",
    year = "2015"
}

@article{Deser:1993yx,
    author = "Deser, Stanley and Schwimmer, A.",
    title = "{Geometric classification of conformal anomalies in arbitrary dimensions}",
    eprint = "hep-th/9302047",
    archivePrefix = "arXiv",
    reportNumber = "BRX-343, SISSA-14-93-EP",
    doi = "10.1016/0370-2693(93)90934-A",
    journal = "Phys. Lett. B",
    volume = "309",
    pages = "279--284",
    year = "1993"
}

@article{Cai:2009ua,
    author = "Cai, Rong-Gen and Cao, Li-Ming and Ohta, Nobuyoshi",
    title = "{Black Holes in Gravity with Conformal Anomaly and Logarithmic Term in Black Hole Entropy}",
    eprint = "0911.4379",
    archivePrefix = "arXiv",
    primaryClass = "hep-th",
    reportNumber = "KU-TP-039",
    doi = "10.1007/JHEP04(2010)082",
    journal = "JHEP",
    volume = "04",
    pages = "082",
    year = "2010"
}

@article{Cai:2014jea,
    author = "Cai, Rong-Gen",
    title = "{Thermodynamics of Conformal Anomaly Corrected Black Holes in AdS Space}",
    eprint = "1405.1246",
    archivePrefix = "arXiv",
    primaryClass = "hep-th",
    doi = "10.1016/j.physletb.2014.04.044",
    journal = "Phys. Lett. B",
    volume = "733",
    pages = "183--189",
    year = "2014"
}

@article{Fernandes:2023vux,
    author = "Fernandes, Pedro G. S.",
    title = "{Rotating black holes in semiclassical gravity}",
    eprint = "2305.10382",
    archivePrefix = "arXiv",
    primaryClass = "gr-qc",
    doi = "10.1103/PhysRevD.108.L061502",
    journal = "Phys. Rev. D",
    volume = "108",
    number = "6",
    pages = "L061502",
    year = "2023"
}

@article{Gurses:2023ahu,
    author = "Gurses, Metin and Tekin, Bayram",
    title = "{Kerr-Vaidya type radiating black holes in semiclassical gravity with conformal anomaly}",
    eprint = "2310.00312",
    archivePrefix = "arXiv",
    primaryClass = "gr-qc",
    doi = "10.1103/PhysRevD.109.024001",
    journal = "Phys. Rev. D",
    volume = "109",
    number = "2",
    pages = "024001",
    year = "2024"
}

@article{Solodukhin:1997yy,
    author = "Solodukhin, Sergey N.",
    title = "{Entropy of Schwarzschild black hole and string - black hole correspondence}",
    eprint = "hep-th/9701106",
    archivePrefix = "arXiv",
    reportNumber = "WATPHYS-TH-97-01",
    doi = "10.1103/PhysRevD.57.2410",
    journal = "Phys. Rev. D",
    volume = "57",
    pages = "2410--2414",
    year = "1998"
}

@article{Kaul:2000kf,
    author = "Kaul, Romesh K. and Majumdar, Parthasarathi",
    title = "{Logarithmic correction to the Bekenstein-Hawking entropy}",
    eprint = "gr-qc/0002040",
    archivePrefix = "arXiv",
    doi = "10.1103/PhysRevLett.84.5255",
    journal = "Phys. Rev. Lett.",
    volume = "84",
    pages = "5255--5257",
    year = "2000"
}

@article{Das:2001ic,
    author = "Das, Saurya and Majumdar, Parthasarathi and Bhaduri, Rajat K.",
    title = "{General logarithmic corrections to black hole entropy}",
    eprint = "hep-th/0111001",
    archivePrefix = "arXiv",
    doi = "10.1088/0264-9381/19/9/302",
    journal = "Class. Quant. Grav.",
    volume = "19",
    pages = "2355--2368",
    year = "2002"
}

@article{Dolan:2014vba,
    author = "Dolan, Brian P. and Kostouki, Anna and Kubiznak, David and Mann, Robert B.",
    title = "{Isolated critical point from Lovelock gravity}",
    eprint = "1407.4783",
    archivePrefix = "arXiv",
    primaryClass = "hep-th",
    doi = "10.1088/0264-9381/31/24/242001",
    journal = "Class. Quant. Grav.",
    volume = "31",
    number = "24",
    pages = "242001",
    year = "2014"
}

@article{Hu:2024ldp,
    author = "Hu, Ya-Peng and An, Yu-Sen and Sun, Gao-Yong and You, Wen-Long and Shi, Da-Ning and Zhang, Hongsheng and Chen, Xiaosong and Cai, Rong-Gen",
    title = "{Quantum anomaly triggers the violation of scaling laws in gravitational system}",
    eprint = "2410.23783",
    archivePrefix = "arXiv",
    primaryClass = "gr-qc",
    month = "10",
    year = "2024"
}

@article{Zhang:2023bzv,
    author = "Zhang, Zhenyu and Hou, Yehui and Guo, Minyong",
    title = "{Observational signatures of rotating black holes in the semiclassical gravity with trace anomaly*}",
    eprint = "2305.14924",
    archivePrefix = "arXiv",
    primaryClass = "gr-qc",
    doi = "10.1088/1674-1137/ad432b",
    journal = "Chin. Phys. C",
    volume = "48",
    number = "8",
    pages = "085106",
    year = "2024"
}

@article{Das:2025eiv,
    author = "Das, Surajit and Dalui, Surojit and Lee, Bum-Hoon and Cai, Yi-Fu",
    title = "{Extreme-Mass-Ratio Inspirals Embedded in Dark Matter Halo II: Chaotic Imprints in Gravitational Waves}",
    eprint = "2512.04848",
    archivePrefix = "arXiv",
    primaryClass = "gr-qc",
    month = "12",
    year = "2025"
}

@article{Huang:2020rjf,
    author = "Huang, Shun-Jia and Hu, Yi-Ming and Korol, Valeriya and Li, Peng-Cheng and Liang, Zheng-Cheng and Lu, Yang and Wang, Hai-Tian and Yu, Shenghua and Mei, Jianwei",
    title = "{Science with the TianQin Observatory: Preliminary results on Galactic double white dwarf binaries}",
    eprint = "2005.07889",
    archivePrefix = "arXiv",
    primaryClass = "astro-ph.HE",
    doi = "10.1103/PhysRevD.102.063021",
    journal = "Phys. Rev. D",
    volume = "102",
    number = "6",
    pages = "063021",
    year = "2020"
}

@article{Korol:2017qcx,
    author = "Korol, Valeriya and Rossi, Elena M. and Groot, Paul J. and Nelemans, Gijs and Toonen, Silvia and Brown, Anthony G. A.",
    title = "{Prospects for detection of detached double white dwarf binaries with Gaia, LSST and LISA}",
    eprint = "1703.02555",
    archivePrefix = "arXiv",
    primaryClass = "astro-ph.HE",
    doi = "10.1093/mnras/stx1285",
    journal = "Mon. Not. Roy. Astron. Soc.",
    volume = "470",
    number = "2",
    pages = "1894--1910",
    year = "2017"
}

@article{Klein:2015hvg,
    author = "Klein, Antoine and others",
    title = "{Science with the space-based interferometer eLISA: Supermassive black hole binaries}",
    eprint = "1511.05581",
    archivePrefix = "arXiv",
    primaryClass = "gr-qc",
    doi = "10.1103/PhysRevD.93.024003",
    journal = "Phys. Rev. D",
    volume = "93",
    number = "2",
    pages = "024003",
    year = "2016"
}

@article{Wang:2019ryf,
    author = "Wang, Hai-Tian and others",
    title = "{Science with the TianQin observatory: Preliminary results on massive black hole binaries}",
    eprint = "1902.04423",
    archivePrefix = "arXiv",
    primaryClass = "astro-ph.HE",
    doi = "10.1103/PhysRevD.100.043003",
    journal = "Phys. Rev. D",
    volume = "100",
    number = "4",
    pages = "043003",
    year = "2019"
}

@article{Sesana:2016ljz,
    author = "Sesana, Alberto",
    title = "{Prospects for Multiband Gravitational-Wave Astronomy after GW150914}",
    eprint = "1602.06951",
    archivePrefix = "arXiv",
    primaryClass = "gr-qc",
    doi = "10.1103/PhysRevLett.116.231102",
    journal = "Phys. Rev. Lett.",
    volume = "116",
    number = "23",
    pages = "231102",
    year = "2016"
}

@article{Liu:2020eko,
    author = "Liu, Shuai and Hu, Yi-Ming and Zhang, Jian-dong and Mei, Jianwei",
    title = "{Science with the TianQin observatory: Preliminary results on stellar-mass binary black holes}",
    eprint = "2004.14242",
    archivePrefix = "arXiv",
    primaryClass = "astro-ph.HE",
    doi = "10.1103/PhysRevD.101.103027",
    journal = "Phys. Rev. D",
    volume = "101",
    number = "10",
    pages = "103027",
    year = "2020"
}

@article{Babak:2017tow,
    author = "Babak, Stanislav and Gair, Jonathan and Sesana, Alberto and Barausse, Enrico and Sopuerta, Carlos F. and Berry, Christopher P. L. and Berti, Emanuele and Amaro-Seoane, Pau and Petiteau, Antoine and Klein, Antoine",
    title = "{Science with the space-based interferometer LISA. V: Extreme mass-ratio inspirals}",
    eprint = "1703.09722",
    archivePrefix = "arXiv",
    primaryClass = "gr-qc",
    doi = "10.1103/PhysRevD.95.103012",
    journal = "Phys. Rev. D",
    volume = "95",
    number = "10",
    pages = "103012",
    year = "2017"
}

@article{Fan:2020zhy,
    author = "Fan, Hui-Min and Hu, Yi-Ming and Barausse, Enrico and Sesana, Alberto and Zhang, Jian-dong and Zhang, Xuefeng and Zi, Tie-Guang and Mei, Jianwei",
    title = "{Science with the TianQin observatory: Preliminary result on extreme-mass-ratio inspirals}",
    eprint = "2005.08212",
    archivePrefix = "arXiv",
    primaryClass = "astro-ph.HE",
    doi = "10.1103/PhysRevD.102.063016",
    journal = "Phys. Rev. D",
    volume = "102",
    number = "6",
    pages = "063016",
    year = "2020"
}

@article{TianQin:2020hid,
    author = "Mei, Jianwei and others",
    collaboration = "TianQin",
    title = "{The TianQin project: current progress on science and technology}",
    eprint = "2008.10332",
    archivePrefix = "arXiv",
    primaryClass = "gr-qc",
    doi = "10.1093/ptep/ptaa114",
    journal = "PTEP",
    volume = "2021",
    number = "5",
    pages = "05A107",
    year = "2021"
}

@article{LISA:2017pwj,
    author = "Amaro-Seoane, Pau and others",
    collaboration = "LISA",
    title = "{Laser Interferometer Space Antenna}",
    eprint = "1702.00786",
    archivePrefix = "arXiv",
    primaryClass = "astro-ph.IM",
    month = "2",
    year = "2017"
}

@article{Hu:2017mde,
    author = "Hu, Wen-Rui and Wu, Yue-Liang",
    title = "{The Taiji Program in Space for gravitational wave physics and the nature of gravity}",
    doi = "10.1093/nsr/nwx116",
    journal = "Natl. Sci. Rev.",
    volume = "4",
    number = "5",
    pages = "685--686",
    year = "2017"
}

@article{Yunes:2011aa,
    author = "Yunes, Nicolas and Pani, Paolo and Cardoso, Vitor",
    title = "{Gravitational Waves from Quasicircular Extreme Mass-Ratio Inspirals as Probes of Scalar-Tensor Theories}",
    eprint = "1112.3351",
    archivePrefix = "arXiv",
    primaryClass = "gr-qc",
    doi = "10.1103/PhysRevD.85.102003",
    journal = "Phys. Rev. D",
    volume = "85",
    pages = "102003",
    year = "2012"
}

@article{Canizares:2012is,
    author = "Canizares, Priscilla and Gair, Jonathan R. and Sopuerta, Carlos F.",
    title = "{Testing Chern-Simons Modified Gravity with Gravitational-Wave Detections of Extreme-Mass-Ratio Binaries}",
    eprint = "1205.1253",
    archivePrefix = "arXiv",
    primaryClass = "gr-qc",
    doi = "10.1103/PhysRevD.86.044010",
    journal = "Phys. Rev. D",
    volume = "86",
    pages = "044010",
    year = "2012"
}

@article{Moore:2017lxy,
    author = "Moore, Christopher J. and Chua, Alvin J. K. and Gair, Jonathan R.",
    title = "{Gravitational waves from extreme mass ratio inspirals around bumpy black holes}",
    eprint = "1707.00712",
    archivePrefix = "arXiv",
    primaryClass = "gr-qc",
    doi = "10.1088/1361-6382/aa85fa",
    journal = "Class. Quant. Grav.",
    volume = "34",
    number = "19",
    pages = "195009",
    year = "2017"
}

@article{Martel:2000rn,
    author = "Martel, Karl and Poisson, Eric",
    title = "{Regular coordinate systems for Schwarzschild and other spherical space-times}",
    eprint = "gr-qc/0001069",
    archivePrefix = "arXiv",
    doi = "10.1119/1.1336836",
    journal = "Am. J. Phys.",
    volume = "69",
    pages = "476--480",
    year = "2001"
}

@book{Baumgarte:2010ndz,
    author = "Baumgarte, Thomas W. and Shapiro, Stuart L.",
    title = "{Numerical Relativity: Solving Einstein's Equations on the Computer}",
    doi = "10.1017/CBO9781139193344",
    publisher = "Cambridge University Press",
    year = "2010"
}

@article{Grossman:2011im,
    author = "Grossman, Rebecca and Levin, Janna and Perez-Giz, Gabe",
    title = "{Faster computation of adiabatic extreme mass-ratio inspirals using resonances}",
    eprint = "1108.1819",
    archivePrefix = "arXiv",
    primaryClass = "gr-qc",
    doi = "10.1103/PhysRevD.88.023002",
    journal = "Phys. Rev. D",
    volume = "88",
    number = "2",
    pages = "023002",
    year = "2013"
}

@article{Zi:2023qfk,
    author = "Zi, Tieguang and Li, Peng-Cheng",
    title = "{Gravitational waves from extreme-mass-ratio inspirals in the semiclassical gravity spacetime}",
    eprint = "2311.07279",
    archivePrefix = "arXiv",
    primaryClass = "gr-qc",
    doi = "10.1103/PhysRevD.109.064089",
    journal = "Phys. Rev. D",
    volume = "109",
    number = "6",
    pages = "064089",
    year = "2024"
}

@article{Sundararajan:2007jg,
    author = "Sundararajan, Pranesh A. and Khanna, Gaurav and Hughes, Scott A.",
    title = "{Towards adiabatic waveforms for inspiral into Kerr black holes. I. A New model of the source for the time domain perturbation equation}",
    eprint = "gr-qc/0703028",
    archivePrefix = "arXiv",
    doi = "10.1103/PhysRevD.76.104005",
    journal = "Phys. Rev. D",
    volume = "76",
    pages = "104005",
    year = "2007"
}

@article{Babak:2006uv,
    author = "Babak, Stanislav and Fang, Hua and Gair, Jonathan R. and Glampedakis, Kostas and Hughes, Scott A.",
    title = "{'Kludge' gravitational waveforms for a test-body orbiting a Kerr black hole}",
    eprint = "gr-qc/0607007",
    archivePrefix = "arXiv",
    doi = "10.1103/PhysRevD.75.024005",
    journal = "Phys. Rev. D",
    volume = "75",
    pages = "024005",
    year = "2007",
    note = "[Erratum: Phys.Rev.D 77, 04990 (2008)]"
}

@article{https://doi.org/10.1002/zamm.19630430611,
author = {Macke, W.},
title = {L. D. Landau and E. M. Lifshitz, The Classical Theory of Fields. Revised second edition. Translated from the Russian. IX + 404 S. m. 19 Abb. Oxford/London/Paris/Frankfurt 1962. Pergamon Press. Preis geb. 80s. net},
journal = {ZAMM - Journal of Applied Mathematics and Mechanics / Zeitschrift für Angewandte Mathematik und Mechanik},
volume = {43},
number = {6},
pages = {287-287},
doi = {https://doi.org/10.1002/zamm.19630430611},
url = {https://onlinelibrary.wiley.com/doi/abs/10.1002/zamm.19630430611},
eprint = {https://onlinelibrary.wiley.com/doi/pdf/10.1002/zamm.19630430611},
year = {1963}
}

@article{Thorne:1980ru,
    author = "Thorne, K. S.",
    title = "{Multipole Expansions of Gravitational Radiation}",
    doi = "10.1103/RevModPhys.52.299",
    journal = "Rev. Mod. Phys.",
    volume = "52",
    pages = "299--339",
    year = "1980"
}

@article{Press:1977ps,
    author = "Press, W. H.",
    title = "{Gravitational Radiation from Sources Which Extend Into their Own Wave Zone}",
    doi = "10.1103/PhysRevD.15.965",
    journal = "Phys. Rev. D",
    volume = "15",
    pages = "965--968",
    year = "1977"
}

@book{Poisson_Will_2014, 
place={Cambridge}, 
title={Gravity: Newtonian, Post-Newtonian, Relativistic}, 
publisher={Cambridge University Press}, author={Poisson, Eric and Will, Clifford M.}, 
year={2014}
}

@article{Yang:2024lmj,
    author = "Yang, Sen and Zhang, Yu-Peng and Zhu, Tao and Zhao, Li and Liu, Yu-Xiao",
    title = "{Gravitational waveforms from periodic orbits around a quantum-corrected black hole}",
    eprint = "2407.00283",
    archivePrefix = "arXiv",
    primaryClass = "gr-qc",
    doi = "10.1088/1475-7516/2025/01/091",
    journal = "JCAP",
    volume = "01",
    pages = "091",
    year = "2025"
}

@bOOK{1992rcd..book.....L,
       author = {{Lichtenberg}, A. and {Lieberman}, M.},
        title = "{Regular and Chaotic Dynamics}",
         year = 1992,
       adsurl = {https://ui.adsabs.harvard.edu/abs/1992rcd..book.....L}
}

@article{Suzuki:1999si,
    author = "Suzuki, Shingo and Maeda, Kei-ichi",
    title = "{Signature of chaos in gravitational waves from a spinning particle}",
    eprint = "gr-qc/9910064",
    archivePrefix = "arXiv",
    reportNumber = "WU-AP-83-99",
    doi = "10.1103/PhysRevD.61.024005",
    journal = "Phys. Rev. D",
    volume = "61",
    pages = "024005",
    year = "2000"
}

@article{Kiuchi:2004bv,
    author = "Kiuchi, Kenta and Maeda, Kei-ichi",
    title = "{Gravitational waves from chaotic dynamical system}",
    eprint = "gr-qc/0404124",
    archivePrefix = "arXiv",
    reportNumber = "WU-AP-185-04",
    doi = "10.1103/PhysRevD.70.064036",
    journal = "Phys. Rev. D",
    volume = "70",
    pages = "064036",
    year = "2004"
}

@article{Drasco:2003ky,
    author = "Drasco, Steve and Hughes, Scott A.",
    title = "{Rotating black hole orbit functionals in the frequency domain}",
    eprint = "astro-ph/0308479",
    archivePrefix = "arXiv",
    reportNumber = "CSR-03-51",
    doi = "10.1103/PhysRevD.69.044015",
    journal = "Phys. Rev. D",
    volume = "69",
    pages = "044015",
    year = "2004"
}

@article{Robson:2018ifk,
    author = "Robson, Travis and Cornish, Neil J. and Liu, Chang",
    title = "{The construction and use of LISA sensitivity curves}",
    eprint = "1803.01944",
    archivePrefix = "arXiv",
    primaryClass = "astro-ph.HE",
    doi = "10.1088/1361-6382/ab1101",
    journal = "Class. Quant. Grav.",
    volume = "36",
    number = "10",
    pages = "105011",
    year = "2019"
}

@article{Li:2024rnk,
    author = "Li, En-Kun and others",
    title = "{Gravitational wave astronomy with TianQin}",
    eprint = "2409.19665",
    archivePrefix = "arXiv",
    primaryClass = "astro-ph.GA",
    doi = "10.1088/1361-6633/adc9be",
    journal = "Rept. Prog. Phys.",
    volume = "88",
    number = "5",
    pages = "056901",
    year = "2025"
}

@article{Liu:2023qap,
    author = "Liu, Chang and Ruan, Wen-Hong and Guo, Zong-Kuan",
    title = "{Confusion noise from Galactic binaries for Taiji}",
    eprint = "2301.02821",
    archivePrefix = "arXiv",
    primaryClass = "astro-ph.IM",
    doi = "10.1103/PhysRevD.107.064021",
    journal = "Phys. Rev. D",
    volume = "107",
    number = "6",
    pages = "064021",
    year = "2023"
}

\end{document}